\documentclass[english,floatfix,twocolumn,showpacs,preprintnumbers,amsmath,amssymb]{revtex4}
\usepackage{array}
\usepackage{graphicx}
\usepackage{amssymb}
\usepackage{longtable}

\makeatletter

\providecommand{\tabularnewline}{\\}

\usepackage{bm}

\usepackage{babel}
\makeatother
\begin{document}

\title{Parameterization of the Woods-Saxon Potential for Shell-Model Calculations}

\author{N.Schwierz }

\altaffiliation[Also at ]{Physics Department, Universit\"at Konstanz, 
Konstanz, Germany }

\author{I. Wiedenh\"over}

\author{A. Volya}

\affiliation{Florida State University, Physics Department, Tallahassee, Fl 32306 }

\date{\today}

\begin{abstract}
The drastically expanded use of the Woods-Saxon potential in modern
day nuclear physics and the availability of new nuclear data motivated
us to review and optimize the parameters of this potential to the
experimental single-nucleon spectra around the doubly-magic nuclei
between $^{16}$O and $^{208}$Pb. We obtain a parameterization which
is applicable over the whole nuclear chart for nuclides between $^{16}$O
and the heaviest elements. Apart from Coulomb components the obtained
parameter set is isospin symmetric. We demonstrate that the potential
provides a good description of the nuclear mean field leading to quality
single-particle spectra, nuclear radii, prediction of drip-lines,
shell closures and other properties. This presented Woods-Saxon fit
provides an adequate single-particle basis for shell model calculations
bridging over into the continuum. 
\end{abstract}

\maketitle

\section{\label{sec:Intro}Introduction}

The description of the nuclear many-body system by an effective mean
field is a doorway to the understanding of atomic nuclei. The averaged
single nucleon dynamics in the field of all other nucleons is a starting
point in practically all many-body methods. In this context, a good
choice of single-nucleon basis states is the key for success in any
quantum many-body approach. A plethora of experimental observations
such as magic numbers, shell gaps, binding energies, nuclear radii,
abundances of nuclei in nature and reaction properties all confirm
the remarkable success of the very simple, pure mean field picture.

Generally, the motion of non-interacting particles in the mean field
is not an exact solution to the many-body problem. Residual interactions
or collective dynamics of the mean field itself are present. The Hartree-Fock
or Hartree-Fock-Bogoliubov approach allows to variationally find the
best possible mean field thus minimizing the residual interactions.
The technique has been demonstrated to be very successful. Nevertheless,
at present it is still a challenge to find the best mean field with
properly preserved symmetries, not to mention the interactions and
related physics that go beyond the Hartree-Fock approach.

In contrast, in a very simple approach the mean field can be taken
in the form of a three dimensional harmonic oscillator, which provides
an analytical set of basis states. The possibility of an exact translationally
invariant treatment with full center of mass extraction is particularly
appealing. Historically, the harmonic oscillator mean field with added
spin orbit term was the first successful mean field treatment, by
which the correct sequence of orbitals and the magic numbers was predicted
\cite{Goe49}.

In order to achieve a quantitative description of nuclei, oscillator
parameters have to be selected in dependence of the nuclear mass.
Typically, for spherical nuclei of mass $A$, the oscillator parameter
$\hbar\omega=41/A^{1/3}$ MeV provides a good description, which was
found by comparing the oscillator rms mass radius with experiment.
The analytical form of the single particle basis comes at the expense
of relatively strong residual particle-particle interactions. Still,
very successful shell-model descriptions have been developed on the
basis of the harmonic oscillator \cite{brown}, where the many-body
Hamiltonian that includes residual interactions is diagonalized exactly.

Present day nuclear physics has expanded its reach far beyond the
valley of stability into regions of the nuclear chart where the
continuum spectrum of the mean field potential becomes important. To
address these situations or for the purposes of nuclear reaction
physics, the choice of a Woods-Saxon \cite{WS54} potential as a model
for the mean field has become a common approach. The analyticity of
solution in the harmonic oscillator is traded for the presence of a
continuum spectrum and a potential resembling the geometric
distribution of nuclear density. Although the single particle states
must be found numerically, such computations are trivial with today's
computers, making the numerical approach preferential over
analytically solvable models such as Ginocchio potential
\cite{ginocchio85} or the square well. Other forms of potentials have
been considered in the past \cite{feshbach58}.  With little exception,
all modern theoretical techniques dealing with the physics on the
interface of structure and reactions have their roots in the
Woods-Saxon potential \cite{GSM,proton}.

A number of parameterizations of the Woods-Saxon potential have been
published, created with different objectives and relevant to different
nuclear mass regions. Most commonly used is the so-called
{}``Universal'' parameterization \cite{Dud82} , which was adjusted to
reproduce the single-particle binding energies of proton and neutron
orbitals around the doubly-magic nucleus $^{208}$Pb and correct ground
state spins for nuclei of masses around A=180, but claimed to be
applicable to lighter mass regions as well. Characteristic for the
{}``Universal'' parameterization is the choice of different radii for
the proton- and neutron potentials.  It has been pointed out, that
this parameterization has shortcomings with respect to lighter nuclei
and that it predicts charge radii inconsistent with
experiment. \cite{Dud95}.

The importance of a good starting parameterization of the mean field
potential, its expanded use in the context of the continuum shell
model and the availability of new experimental data motivated us to
revisit the question of parameters in the Woods-Saxon potential. In
this work we find a set of parameters to be considered as a global
Woods-Saxon parameterization, obeying the symmetry principles of isospin
conservation in the nuclear interaction as well as correct two-body
kinematics. We use a set of single particle and single hole states
in the vicinity of doubly-magic nuclei $^{16}$O, $^{40}$Ca, $^{48}$Ca,
$^{56}$Ni, $^{100}$Sn, $^{132}$Sn and $^{208}$Pb to provide experimental
input on the single particle data. We adjust the potential parameters
to reproduce the experimental data by applying a least-squares fit
algorithm.

This work is a continuation and extension of a previous effort to
determine a globally applicable set of Woods-Saxon potential parameters.
Among the similarities between our work and the previous parameter
set is the isospin-symmetry of the nuclear potential. While the previously
obtained parameter set was not published, a total-Routhian-surface
calculation based on it was discussed in \cite{Ced94}.

The paper is organized as follows. In section \ref{sec:SPdata}, we
describe the experimental data that is used in optimizing the
parameters, in section \ref{sec:WSCalc} we address the properties of
the Woods-Saxon Hamiltonian, its symmetries and the
parameterization. In section \ref{sec:fit} we describe the calculation
of the orbitals of the Woods-Saxon potential and the least squares fit
algorithm employed to adjust the parameters to fit the experimental
data. The predictions of our Woods-Saxon potential for rms charge
radii and neutron radii is discussed in section \ref{sec:Results}, as
well as its predictions for certain neutron-rich nuclides.

\section{\label{sec:SPdata}Single Particle Level Data}

Basis for this investigation are the experimental single particle
spectra around the doubly magic nuclei $^{16}$O, $^{40}$Ca, $^{48}$Ca,
$^{56}$Ni, $^{132}$Sn and $^{208}$Pb. These systems were chosen,
because they are expected to show the most pure experimental manifestation
of single-particle excitations. The binding energies of the single-nucleon
orbitals around a magic nucleus of mass A were extracted from the
excited states of the neighboring nuclei by

\begin{equation}
\epsilon(A,I^{\pi})=\Delta M(A)-(\Delta M(A-1)+\Delta M(1))-E(I^{\pi})\label{she}\end{equation}
 for hole-states, \begin{equation}
\epsilon(A+1,I^{\pi})=\Delta M(A+1)-(\Delta M(A)+\Delta M(1))+E(I^{\pi})\label{spe}\end{equation}
 for particle-states. Here $\Delta M(A)$ denotes the experimental
mass defects of a nucleus with mass number $A$ and $E(I^{\pi})$
the excitation energy of a given state. Both proton and neutron orbitals
are considered.

The experimental data were generally extracted from the NNDC data base
\cite{A=15} - \cite{A=209}, selecting data sets representing
single-proton or neutron transfer or pickup reactions, where
available.  For the $^{40}$Ca, $^{48}$Ca and $^{208}$Pb nuclides, some
of the single-particle and single-hole strength was observed to be
fragmented over a number of individual states. In these cases, we
calculate the orbital energy from an average excitation energy of the
states, weighted with the single-particle spectroscopic factors. The
cases for which the data base showed contradictory experimental
results are discussed separately in the following paragraphs.  The
data set of orbital energies used in this study and the relevant
experimental information is summarized in the Table
\ref{tab:orbene16O}.

In the case of $^{17}$F and $^{17}$O, we used $d_{5/2}$ and $s_{1/2}$
states established by direct transfer-reactions. However, the
corresponding $d_{3/2}$ orbits are unbound and their strength is
fragmented over a number of resonances, both in $^{17}$F and
$^{17}$O. The resonance data for neutron scattering on $^{16}$O was
taken from Ref. \cite{Johnson}, including the analysis of
spectroscopic factors.  The analogous factors for protons were
extracted from the ratio of the observed proton-resonance widths
\cite{A=17} to a single-particle decay width calculated with our
Woods-Saxon potential for the experimental kinematics. The results
are summarized in Tab. \ref{A17tab}.

\begin{table}
\begin{tabular}{|c|c|c|c||c|c|c|c|}
\hline 
\multicolumn{4}{|c||}{1d$_{3/2}$ neutron $^{17}$O} & \multicolumn{4}{c|}{1d$_{3/2}$ proton $^{17}$F}\tabularnewline
\hline
\hline 
Ex & E$_{\nu}$ & $\Gamma_{n}${[}keV] & $S$$^{a}$ & Ex  & E$_{\pi}$ & $\Gamma_{p}${[}keV] & $S$$^{b}$\tabularnewline
\hline
\hline 
5.085 & 0.937 & 96 & 0.689 & 5.000 & 4.400 & 1530 & 0.845\tabularnewline
\hline 
5.869 & 1.721 & 6.6 & 0.010 & 5.820 & 5.220 & 180 & 0.057\tabularnewline
\hline 
7.202 & 3.052 & 280 & 0.169 & 7.479 & 6.879 & 795 & 0.097\tabularnewline
\hline 
8.070 & 3.916 & 71 & 0.085 & 8.760 & 8.160 & 90$^{c}$ & 0.000\tabularnewline
\hline 
8.897 & 4.751 & 68 & 0.042 &  &  &  & \tabularnewline
\hline
\hline 
5.836 & 1.710 &  & 0.995 & 5.288  & 4.688 &  & 1.000\tabularnewline
\hline
\end{tabular}\caption{$d_{3/2}$ resonances in $^{17}$O and $^{17}$F, 
energies are in MeV. 
The bottom line corresponds to the adopted
weighted single particle energies. \\ 
$^{a}$The spectroscopic factors
are from R-matrix analysis in Ref. \cite{Johnson}\\
$^{b}$Our parametrization with adjusted potential depth was used
to compute the single particle decay width at the experimental energies,
the ratio of experimental width to single particle determined the
spectroscopic factor $S$. The total strength was normalized to 1.\\
$^{c}$The proton channel is only 20\% of the total width.\label{A17tab}}

\end{table}

For the case of proton orbitals in $^{40}$Ca, the Nuclear Data Sheets
entry for $^{39}$K \cite{A=39} shows a state at 8.43 MeV excitation
energy, which is assigned d$_{3/2}$ character with a spectroscopic
factor of $C^{2}S$=0.24, based on data published in Ref.~\cite{Dev81}.
We excluded this state from our analysis, for two reasons; First it
would represent a rather unusual fragmentation of single-particle
strength, placing a fragment of the ground state at over 8 MeV excitation
energy. Second, including it would lead to inconsistent Coulomb shifts
between the proton and neutron orbitals. The assignment of d$_{3/2}$
in Ref.~\cite{Dev81} was based on an angular distribution measurement,
but the difference between a d$_{3/2}$ assignment and the more likely
d$_{5/2}$ assignment seems inconclusive from the data presented in
the publication. In view of this assessment, we based our analysis
of the $^{39}$K data on the older measurement of \cite{Dol76}. Note
that, after excluding the above mentioned d$_{3/2}$ assignment, the
data from both references \cite{Dol76} and \cite{Dev81} leads to
almost identical energies of s.p.\  orbitals.

For $^{56}$Ni, only one experiment for a direct neutron-transfer
is published, based on the $(d,p)$ reaction \cite{Reh98}. This experiment
revealed spectroscopic factors of single-particle character for the
lowest $3/2^{-},5/2^{-}$ and $1/2^{-}$ states. For the proton orbitals,
no direct spectroscopic factor measurements are available to date,
however, the lowest $3/2^{-},5/2^{-}$ and $1/2^{-}$ states show
consistent Coulomb-shifts, and therefore were assumed to be the analog
proton excitations.

$^{100}$Sn is still beyond the reach of today's radioactive beam
facilities, but systematic studies of the shell structure of nuclides
approaching it have produced a set of effective single particle
orbitals\cite{Gra95}, which we include in our investigation. 

For the doubly-magic nucleus $^{132}$Sn no single-particle reactions
were published to date. However, the level schemes of nuclei around
$^{132}$Sn are well established from $\beta$--decay measurements.
Using data from the NNDC data base \cite{A=131,A=133}, we identified
the lowest states of given spin and parity with the single-particle
or single hole states of the orbitals in question, and experimental
binding energies for the ground states.

$^{208}$Pb has been studied by a multitude of experiments. The
spectrum of single-particle and hole excitations seems well
established by the measured large spectroscopic factors
\cite{A=207,A=209}. Nevertheless, even for this text-book doubly magic
nucleus, some fragmentation of the orbital strengths can be expected
for high excitation energies.  Data on the fragmentation of proton and
neutron hole states around $^{208}$Pb was established in
Refs. \cite{Gal78} and \cite{Gra92}.  Although these experiments show
the main orbital strength to be concentrated in one state for each
orbital, the smaller fragments lead to significant centroid energy
shifts, approximately 0.5 MeV for the deepest hole states. In contrast
to this situation, no data on the fragmentation for the neutron and
proton particle states was found in the literature, in spite of the
large number of published experimental studies. It seems worth while
to revisit experiments on the fragmentation of single particle
structure around $^{208}$Pb with modern experimental techniques in
order to address this apparent deficiency of the experimental data.

\setlength{\LTcapwidth}{\linewidth}
\begin{longtable}{|l|l|l|l|p{3.7cm}|}

\caption[Table I]{Systematics of nuclear single particle energies. 
In the first three columns,
we list the orbital type, extracted binding energy, and excitation
energy of the single particle level. The binding and excitation energies
are weighted with spectroscopic factors (see text), the value of the
total spectroscopic strength ${\cal S}=\sum S=\sum C^{2}S/(2j+1)$
appears in the fourth column for the cases where relevant experimental
information is available. The fifth column identifies individual excited
states used in the analysis.}
 \label{tab:orbene16O}
\tabularnewline
\hline
Orbital &
Energy &
E$_{X}$&
${\cal S}$&
individual states\tabularnewline
&
{[}MeV] &
{[}MeV] &
&
{[}MeV] \tabularnewline
\hline \hline
\endfirsthead

\multicolumn{5}{l}{Table II -- Continued}\tabularnewline
\hline 
Orbital &
Energy &
E$_{X}$&
${\cal S}$&
individual states\tabularnewline
&
{[}MeV] &
{[}MeV] &
&
{[}MeV] \tabularnewline
\hline \hline
\endhead

\endfoot

\hline \hline
\caption*{$^a$ The data was extracted by systematic shell model 
studies of nuclei close to $^{100}$Sn \cite{Gra95}. \\
$^b$ The data is based on the
excitation energy of the lowest state of given spin and parity, no
information on spectroscopic factors is available.
}
\endlastfoot

\multicolumn{5}{|l|}{\textbf{$^{16}$O}}\tabularnewline
\hline 
\multicolumn{3}{|l|}{\textbf{neutron hole}}&
&
$^{15}$O\cite{A=15}\tabularnewline
\hline 
$1p_{1/2}$ &
-15.66 &
0 &
&
0\tabularnewline
\hline 
\multicolumn{3}{|l|}{\textbf{neutron particle}}&
&
$^{17}$O\cite{Johnson}\tabularnewline
\hline 
$1d_{5/2}$ &
-4.14 &
0 &
1.00 &
0\tabularnewline
$2s_{1/2}$ &
-3.27 &
0.87 &
1.00 &
0.87\tabularnewline
$1d_{3/2}$ &
1.710 &
5.836 &
1.00 &
\text{ see Tab. I} \tabularnewline
\hline 
\multicolumn{3}{|l|}{\textbf{proton hole}}&
&
$^{15}$N\cite{A=15}\tabularnewline
\hline 
$1p_{1/2}$ &
-12.13&
0 &
1.13 &
0\tabularnewline
\hline 
\multicolumn{3}{|l|}{\textbf{proton particle}}&
&
$^{17}$F\cite{A=17}\tabularnewline
\hline 
$1d_{5/2}$ &
-0.60 &
0 &
0.94 &
0\tabularnewline
$2s_{1/2}$ &
-0.11 &
0.495 &
0.83&
0.495\tabularnewline
$1d_{3/2}$ &
4.688 &
5.288 &
1.00&
\text{ see Tab. I}
 \tabularnewline
\hline
\hline
\pagebreak[1]
\multicolumn{5}{|l|}{\textbf{$^{40}$Ca}}\tabularnewline
\hline
\multicolumn{3}{|l|}{\textbf{neutron hole}}&
&
$^{39}$Ca \cite{Dol76}\tabularnewline
\hline 
$1d_{5/2}$ &
-22.39 &
6.744 &
0.90 &
21 states between 4.320 and 9.500 MeV\tabularnewline
$2s_{1/2}$ &
-18.19 &
2.533 &
0.82 &
2.463, 4.017\tabularnewline
$1d_{3/2}$ &
-15.64 &
0 &
0.94&
0\tabularnewline
\hline 
\multicolumn{3}{|l|}{\textbf{neutron particle}}&
&
$^{41}$Ca \cite{Uoz94}\tabularnewline
\hline 
$1f_{7/2}$ &
-8.36 &
0 &
0.77 &
0 \tabularnewline
$2p_{3/2}$ &
-5.84 &
2.519 &
0.91 &
1.940, 2.461, 3.731, 4.602\tabularnewline
$2p_{1/2}$ &
-4.20 &
4.157 &
0.70 &
3.613, 3.943, 4.109, 4.754\tabularnewline
$1f_{5/2}$ &
-1.56 &
6.801 &
0.95 &
24 states between 4.878 and 9.084\tabularnewline
\hline 
\multicolumn{3}{|l|}{\textbf{proton hole}}&
 &
$^{39}$K \cite{Dol76}\tabularnewline
\hline 
$1d_{5/2}$ &
-15.07 &
6.738 &
0.83 &
16 states between 5.262 and 9.750\tabularnewline
$2s_{1/2}$ &
-10.92 &
2.593 &
0.87 &
2.520, 4.095\tabularnewline
$1d_{3/2}$ &
-8.33 &
0 &
0.93 &
0\tabularnewline
\hline 
\multicolumn{3}{|l|}{\textbf{proton particle}}&
 &
$^{41}$Sc \cite{A=41}\tabularnewline
\hline 
$1f_{7/2}$ &
-1.09 &
0 &
1.12&
0\tabularnewline
$2p_{3/2}$ &
0.69 &
1.784 &
0.94&
1.716, 2.419\tabularnewline
$2p_{1/2}$ &
2.38 &
3.471 &
0.75 &
3.471\tabularnewline
$1f_{5/2}$ &
4.96 &
5.685 &
0.33&
5.709, 3.192, 5.862, 6.470\tabularnewline
\hline
\hline
\pagebreak[1]
\multicolumn{5}{|l|}{\textbf{$^{48}$Ca}}\tabularnewline
\hline 
\multicolumn{3}{|l|}{\textbf{neutron hole}}&
 &
$^{47}$Ca \cite{A=47}\tabularnewline
\hline 
$1d_{5/2}$ &
-15.61 &
5.669 &
0.15 &
4.980, 5.300, 5.450, 6.250, 6.870\tabularnewline
$2s_{1/2}$ &
-12.55 &
2.600 &
0.90 &
2.600\tabularnewline
$1d_{3/2}$ &
-12.53 &
2.580 &
0.90&
2.580\tabularnewline
$1f_{7/2}$ &
-10.00 &
0.050 &
0.85 &
0, 3.300, 3.430\tabularnewline
\hline 
\multicolumn{3}{|l|}{\textbf{neutron particle}}&
 &
$^{49}$Ca \cite{A=49}\tabularnewline
\hline 
$2p_{3/2}$ &
-4.60 &
0.545 &
0.97 &
0, 4.069\tabularnewline
$2p_{1/2}$ &
-2.86 &
2.282 &
1.03 &
2.021, 4.261\tabularnewline
$1f_{5/2}$ &
-1.20 &
3.946 &
0.95 &
3.586, 3.993\tabularnewline
$1g_{9/2}$ &
0.130 &
5.276 &
0.47 &
4.018, 4.887, 5.378, 6.529, 6.753\tabularnewline
\hline 
\multicolumn{3}{|l|}{\textbf{proton hole}}&
 &
$^{47}$K \cite{A=47}\tabularnewline
\hline 
$1d_{5/2}$ &
-21.47 &
5.664 &
0.62 &
3.432, 5.220, 5.465, 6.462, 7.740,8.020,8.530\tabularnewline
$1d_{3/2}$ &
-16.18 &
0.377 &
1.22 &
0.359, 3.930\tabularnewline
$2s_{1/2}$ &
-16.10 &
0.295 &
0.92 &
0, 3.850\tabularnewline
\hline 
\multicolumn{3}{|l|}{\textbf{proton particle}}&
 &
$^{49}$Sc \cite{A=49}\tabularnewline
\hline 
$1f_{7/2}$ &
-9.35 &
0.278 &
0.91 &
0, 3.809 \tabularnewline
$2p_{3/2}$ &
-6.44 &
3.187&
0.54 &
3.085, 6.717\tabularnewline
$2p_{1/2}$ &
-4.64 &
4.984&
0.88 &
4.495, 5.015, 5.663, 6.816\tabularnewline
\hline
\hline
\pagebreak[1]
\multicolumn{5}{|l|}{\textbf{$^{56}$Ni}}\tabularnewline
\hline
\multicolumn{3}{|l|}{\textbf{neutron hole}}&
 &
$^{55}$Ni \cite{A=55}\tabularnewline
\hline 
$1f_{7/2}$ &
-16.64 &
0 &
&
0\tabularnewline
\hline 
\multicolumn{3}{|l|}{\textbf{neutron particle}}&
 &
$^{57}$Ni \cite{Reh98}\tabularnewline
\hline 
$2p_{3/2}$ &
-10.25 &
0 &
0.91 &
0\tabularnewline
$1f_{5/2}$ &
-9.48 &
0.768 &
0.91 &
0.768\tabularnewline
$2p_{1/2}$ &
-9.13 &
1.113 &
0.90 &
1.113\tabularnewline
\hline 
\multicolumn{3}{|l|}{\textbf{proton hole}}&
 &
$^{55}$Co \cite{A=55}\tabularnewline
\hline 
$1f_{7/2}$ &
-7.17 &
0 &
&
0\tabularnewline
\hline 
\multicolumn{3}{|l|}{\textbf{proton particle}}&
&
$^{57}$Cu \cite{A=57}\tabularnewline
\hline 
$2p_{3/2}$ &
-0.69 &
0 &
&
0\tabularnewline
$1f_{5/2}$ &
0.33 &
1.028 &
&
1.028\tabularnewline
$2p_{1/2}$ &
0.41 &
1.106 &
&
1.106\tabularnewline
\hline
\hline
\multicolumn{5}{|l|}{\textbf{$^{100}$Sn} \cite{Gra95}, 
see note$^a$}\tabularnewline
\hline 
\multicolumn{5}{|l|}{\textbf{neutrons} }\tabularnewline
\hline 
$2p_{1/2}$ &
-18.38(20) & & & \tabularnewline
$1g_{9/2}$ &
-17.93(30) & & & \tabularnewline
$2d_{5/2}$ &
-11.13(20) & & & \tabularnewline
$1g_{7/2}$ &
-10.93(20) & & & \tabularnewline
$3s_{1/2}$ &
-9.3(5) & & & \tabularnewline
$1h_{11/2}$ &
-8.6(5) & & & \tabularnewline
$2d_{3/2}$ &
-9.2(5) & & & \tabularnewline
\hline 
\multicolumn{5}{|l|}{\textbf{protons} }\tabularnewline
\hline 
$1f_{5/2}$ &
-8.71 & & & \tabularnewline
$2p_{3/2}$ &
-6.38 & & & \tabularnewline
$2p_{1/2}$ &
-3.53(20) & & & \tabularnewline
$1g_{9/2}$ &
-2.92(20) & & & \tabularnewline
$2d_{5/2}$ &
3.00(80) & & & \tabularnewline
$1g_{7/2}$ &
3.90(15) & & & \tabularnewline
\hline
\hline
\pagebreak[4]
\multicolumn{5}{|l|}{\textbf{$^{132}$Sn}, see note $^b$} \tabularnewline
\hline 
\multicolumn{4}{|l|}{\textbf{neutron hole}}&
$^{131}$Sn \cite{A=131}\tabularnewline
\hline 
$1g_{7/2}$ &
-9.75 &
2.434 & & 2.434\tabularnewline
$2d_{5/2}$ &
-8.97 &
1.655& & 1.655\tabularnewline
$3s_{1/2}$ &
-7.64 &
0.331& & 0.331\tabularnewline
$1h_{11/2}$&
-7.54 &
0.224& & 0.224\tabularnewline
$2d_{3/2}$ &
-7.31 &
0& & 0\tabularnewline
\hline 
\multicolumn{4}{|l|}{\textbf{neutron particle}}&
$^{133}$Sn \cite{A=133}\tabularnewline
\hline 
$2f_{7/2}$ &
-2.47 &
0& &0 \tabularnewline
$3p_{3/2}$ &
-1.57 &
0.854& &0.854 \tabularnewline
$1h_{9/2}$ &
-0.86 &
1.561& &1.561 \tabularnewline
$2f_{5/2}$ &
-0.42 &
2.004& &2.004 \tabularnewline
\hline 
\multicolumn{4}{|l|}{\textbf{proton hole}}&
$^{131}$In \cite{A=131}\tabularnewline
\hline 
$2p_{1/2}$ &
-16.01 &
0.302& & 0.302\tabularnewline
$1g_{9/2}$ &
-15.71 &
0& & 0\tabularnewline
\hline 
\multicolumn{4}{|l|}{\textbf{proton particle}}&
$^{133}$Sb \cite{A=133}\tabularnewline
\hline 
$1g_{7/2}$ &
-9.68 &
0& & 0\tabularnewline
$2d_{5/2}$ &
-8.72 &
0.962& & 0.962\tabularnewline
$2d_{3/2}$ &
-6.97 &
2.708& & 2.708\tabularnewline
$1h_{11/2}$&
-6.89 &
2.793& & 2.793\tabularnewline
\hline
\hline
\pagebreak[1]
\multicolumn{5}{|l|}{\textbf{$^{208}$Pb}}\tabularnewline
\hline 
\multicolumn{3}{|l|}{\textbf{neutron hole}}&
 &
$^{207}$Pb \tabularnewline
\hline 
$1h_{9/2}$ &
-11.40&
4.036&
0.98&
3.400,5.410,5.620 \cite{Gal78} \tabularnewline
$2f_{7/2}$ &
-9.81&
2.439&
0.95&
2,340,4.570 \cite{Gal78} \tabularnewline
$1i_{13/2}$&
-9.24&
1.870&
0.91&
1.630,5.990 \cite{Gal78} \tabularnewline
$3p_{3/2}$ &
-8.26&
0.89 &
0.88&
0.890 \cite{A=207} \tabularnewline
$2f_{5/2}$ &
-7.94&
0.57 &
0.60 &
0.570 \cite{A=207} \tabularnewline
$3p_{1/2}$ &
-7.37&
0 &
0.90 &
0 \cite{A=207} \tabularnewline
\hline 
\multicolumn{3}{|l|}{\textbf{neutron particle}}&
 &
$^{209}$Pb \cite{A=209}\tabularnewline
\hline 
$2g_{9/2}$ &
-3.94&
0&
0.83&
0\tabularnewline
$1i_{11/2}$&
-3.16&
0.779&
0.86&
0.779\tabularnewline
$1j_{15/2}$&
-2.51&
1.424&
0.58&
1.424\tabularnewline
$3d_{5/2}$ &
-2.37&
1.565&
0.98&
1.565\tabularnewline
$4s_{1/2}$ &
-1.90&
2.033&
0.98&
2.033\tabularnewline
$2g_{7/2}$ &
-1.44&
2.492&
1.05&
2.492\tabularnewline
$3d_{3/2}$ &
-1.40&
2.537&
1.09&
2.537\tabularnewline
\hline 
\multicolumn{3}{|l|}{\textbf{proton hole}}&
 &
$^{207}$Tl \cite{Gra92}\tabularnewline
\hline 
$1g_{7/2}$ &
-12.00 &
3.991 &
0.52&
3.469,3.995,4.888\tabularnewline
$2d_{5/2}$ &
-9.82 &
1.806 &
0.65 &
1.683,4.696\tabularnewline
$1h_{11/2}$&
-9.36 &
1.348 &
0.88&
1.348\tabularnewline
$2d_{3/2}$ &
-8.36 &
0.351 &
0.90 &
0.351\tabularnewline
$3s_{1/2}$ &
-8.01 &
0 &
0.85 &
0\tabularnewline
\hline 
\multicolumn{3}{|l|}{\textbf{proton particle}}&
 &
$^{209}$Bi \cite{A=209}\tabularnewline
\hline 
$1h_{9/2}$ &
-3.80 &
0 &
1.00 &
0\tabularnewline
$2f_{7/2}$ &
-2.90 &
0.897 &
1.38 &
0.897 \tabularnewline
$1i_{13/2}$&
-2.10 &
1.697 &
0.93 &
1.612,2.601\tabularnewline
$2f_{5/2}$ &
-0.97 &
2.824 &
0.87 &
2.824\tabularnewline
$3p_{3/2}$ &
-0.68 &
3.116 &
0.98 &
3.116\tabularnewline
$3p_{1/2}$ &
-0.16 &
3.637 &
0.54 &
3.637\tabularnewline

\end{longtable}

\section{\label{sec:WSCalc}Woods-Saxon Calculations}

\subsection{Woods-Saxon Hamiltonian}

In this section, we develop the terms in the Woods-Saxon Hamiltonian
from very general assumptions about the character of the nuclear mean
field. For the most part, we arrive at the conventional Woods-Saxon
Hamiltonian, but the purpose here is to emphasize the aspects of the
mean field description and to emphasize the theoretical foundations
behind the construction. We highlight the kinematic aspects of the
problem and symmetry considerations. The concept of reduced mass and
isospin-symmetry are of particular importance in the parameterization
of the Woods-Saxon Hamiltonian introduced in this work.

The general strategy behind the construction of a center of mass Hamiltonian
for the nuclear mean field starts with the assumption of a scalar
interaction potential between nucleon and core as a sum of the nuclear
and Coulomb parts. We assume the nucleon and a core forming a nucleus
with mass number $A=N+Z$ containing $N$ neutrons and $Z$ protons.
Thus the core has $A'=A-1$ nucleons. In the following, we use the
prime to denote quantum numbers and parameters of the core.

Woods and Saxon \cite{WS54} suggested to model the nuclear mean field
i.e. the nucleon-core interaction with a spherically symmetric potential
that has a Fermi-function form \begin{equation}
f(r,R,a)=\left[1+\exp\left(\frac{r-R}{a}\right)\right]^{-1},\label{ffunc}\end{equation}
 where the size $R$ and diffuseness of the surface $a$ are fixed
parameters of the same units of length as $r$.

The total nuclear potential is defined as \begin{equation}
V(r)=-Vf(r,R,a),\label{central}\end{equation}
 where $V$ represents total strength and the minus sign is introduced
to represent the attractive nature of the interaction.

The electromagnetic force is a second part contributing to the proton-core
interaction. This repulsive potential is fully determined with the
assumption of a given nuclear charge distribution $\rho(r)$. The
solution of the corresponding electrostatics problem gives \begin{equation}
V_{c}(r)=4\pi e\left(\frac{1}{r}\int_{0}^{r}{r'}^{2}\rho(r')dr'+\int_{r}^{\infty}r'\rho(r')dr'\right).\label{coulomb1}\end{equation}
 In the spirit of the Woods-Saxon parameterization it is often assumed
that the nuclear charge distribution is proportional in shape to the
same function (\ref{ffunc}) $\rho(r)\sim f(r,R_{c},a_{c}),$ where
the coefficient of proportionality must be determined from the normalization
of density to the total nuclear charge. The integration in Eq.~(\ref{coulomb1})
along with a normalization of density must be done numerically, which
is often too time consuming. The influence of surface terms on the
strength of the Coulomb interaction is, however, weak. We have numerically
tested, that the diffuseness of the charge distribution can be set
to zero within the precision of the fit discussed below. Furthermore,
for the same reason we have assumed $R_{c}=R$ which removes an extra
unnecessary parameter that has little influence on the outcome. Except
for special cases \cite{proton} these assumptions are typical in
other Woods-Saxon parameterizations. Through this paper we adopt the
following form of the Coulomb potential \begin{equation}
V_{c}(r)={Z'e^{2}}\left\{ \begin{array}{lr}
{(3R^{2}-r^{2})}/({2R^{3}}), & r\le R,\\
1/r, & r>R,\end{array}\label{coulomb}\right.\end{equation}
 which as a result of the above assumptions corresponds to a uniformly
charged sphere of radius $R$, which can be treated analytically.

The understanding of the Woods-Saxon potential as a two-body problem
naturally leads to the introduction of a reduced mass. However, in
most previous treatments, notably \cite{Dud82}, the bare nucleon
masses were used in the kinetic energy term of the Hamiltonian. While
the difference between a reduced or bare nucleon mass is not important
for the bound state spectrum of heavy nuclides, its use would create
serious problems for the description of light nuclides in our calculations,
notably the nucleon orbitals around $^{16}$O. Furthermore, we aim
to apply this potential as a basis for continuum states, where the
kinematic aspect becomes central. Thus, in the kinetic energy operator
we use the following reduced mass \begin{equation}
\mu=\left(\frac{1}{m_{_{\pi}^{\nu}}}+\frac{1}{M'}\right)^{-1},\label{eq:redmass}\end{equation}
 where $m_{_{\pi}^{\nu}}$ is a neutron/proton mass and $M'$ is the
mass of an $A-1$ core, taken to be $(A-1)amu$

In addition to the central and Coulomb potentials identified above,
the relativistic corrections for a nucleon with a Fermi momentum of
typically 200 MeV are important. In order to highlight the nature
of these corrections, we give a non-relativistic reduction of the
two-body Coulomb problem to the order of $(v/c)^{2}$. The correction
to the nuclear potential is discussed below. The Dirac Hamiltonian
reduces to \begin{equation}
H=\frac{{\bf p}^{2}}{2\mu}-\frac{{\bf p}^{4}}{8\mu^{3}}+V_{c}({\bf r})+\frac{1}{4\mu^{2}}\sigma\cdot[\nabla V_{c}({\bf r})\times{\bf p}]+\frac{1}{8\mu^{2}}\nabla^{2}V_{c}({\bf r}).\label{dirac}\end{equation}
 Throughout this section, we use natural (Planck) units where $\hbar=c=1.$
In nuclei the contribution from the the second (kinetic energy) ${\bf p}^{4}$
term is small and therefore ignored. It turns out that, in comparison
to nuclear forces, corrections due to Coulomb given by the last two
terms are also small. For example the last so-called Darwinian term
is constant, and only non-zero inside the uniformly charged sphere.
For a typical nucleus its value $-3Z'e^{2}/(8R^{3}\mu^{2})$ is only
of the order of $20-30$~keV which should be compared with $\sim50$
MeV of a typical total depth $V$ in (\ref{central}). It should also
be stressed that terms of this nature can be subsumed in the phenomenologically
determined parameters of the central potential. The spin-orbit contributions
due to the Coulomb potential in eq. \ref{dirac} are an order of magnitude
smaller than the nuclear contributions and will be ignored.

Based on the previous discussion, only the nuclear spin-orbit contribution
has to be included in descriptions of effective nuclear forces in
addition to $V_{c}$. Eq.~(\ref{dirac}) is not valid as a relativistic
reduction of the nuclear mean field problem due to the complicated
structure of the nucleon-nucleon force. However, symmetry considerations
suggest the same general proportionality of the spin orbit force to
be the gradient of the mean field potential. Thus, the total effective
Hamiltonian becomes \begin{equation}
H=\frac{{\bf p}^{2}}{2\mu}+V(r)+V_{c}(r)+\frac{1}{2\mu^{2}r}\left(\frac{\partial}{\partial r}\tilde{V}(r)\right)\,\,{\bf l\cdot{\bf s},\label{HWS}}\end{equation}
 where -- unlike for the Coulomb field -- the potential $\tilde{V}(r)$
is not equal to the original potential $V(r)$ and may have a different
form factor \cite{greenlees68}. Therefore, the form factor of $\tilde{V}(r)$
is another assumption that goes into construction of the Woods-Saxon
Hamiltonian \begin{equation}
\tilde{V}(r)=\tilde{V}f(r,R_{SO},a_{SO}).\label{eq:vso}\end{equation}
 Here, $R_{SO}$ and $a_{SO}$ stand for the radius and the diffuseness
of the spin-orbit term.

In principle all symmetry preserving forces which involve the single
particle operators ${\bf p}$, ${\bf r}$, ${\bf s}$, ${\bf t}$ and the
core spin and isospin operators ${\bf T}'$ and ${\bf I}'$ can appear
in the particle-core parameterizations. These generally small terms
are extensively discussed and studied within the optical model
approach to reaction physics \cite{hodgson71,hussein73}. The
consideration of an odd-particle mean field that carries quantum
numbers of an unpaired nucleon is beyond the goals of the
parameterization discussed here. Therefore, we assume that the core
generally carries no spin degree of freedom $I'=0$. The only time such
terms can be of relevance is when single-particle energies are to be
extracted from hole states Eq.~(\ref{she}) in this case the particle spin
may couple to identical single-hole quantum numbers of the
core. These terms are generally small and have a $1/A$ dependence
\cite{feshbach58}, just as in Eq.~(\ref{isosp}).  The only
``second-order'' term of concern is a possible isospin-dependence of
the spin-orbit strength i.e. $({\bf T}\cdot {\bf t})({\bf l}\cdot{\bf
s})$ This dependence has been the subject of several studies
\cite{isakov02,isakov04} as it represents one of the most interesting,
fundamental and at the same time controversial questions related to
the nuclear mean field. The example of a non-relativistic reduction of
the Coulomb potential and the more rigorous studies using the Relativistic
Mean Field approach suggest the same sign of isovector term in spin
orbit potential as in the central. However, a substantial amount of
experimental data \cite{isakov02,isakov04, koura00} as well as
theoretical understanding of an issue based on the Walecka model
suggest the opposite. Namely, for the neutron rich nuclei the spin
orbit splitting for the neutrons is larger than for the protons.
Those studies suggest the value $\kappa_{SO} \sim -0.3$ which can be
compared to the typical strength of the isotopic term in the central
potential $\kappa\sim 0.6$ to 0.9; see discussion below for details
and definitions. The more in-depth arguments and understanding of the
spin orbit splitting comes from studies of shell evolution due to the
tensor forces \cite{otsuka05}.  The effects of the $\rho$ meson
exchanges in the nucleon-nucleon interaction are particularly
important for these mean field properties.

We used our data set and fit procedure to assess the variation of
spin-orbit strength with isospin. Unfortunately, as was already
pointed out in Ref. \cite{isakov02,koura00} the data based on
single-particle energies is only weakly sensitive to this parameter.
Within the precision of our study we were unable to distinguish any
significant isospin dependence of the spin orbit term. Thus, with no
better alternative we define the spin orbit strength in our
parameterization with no explicit dependence on isospin.

\subsection{Woods-Saxon parameterization}

The Hamiltonian of the model is defined in Eq.~(\ref{HWS}), where
the central potential is determined in Eq.~(\ref{central}) and the
spin-orbit term is given by Eq.~(\ref{eq:vso}). Both central and
spin-orbit parts have a Woods-Saxon form factor (\ref{ffunc}). The
Coulomb part is given by Eq.~(\ref{coulomb}). For each individual
nucleus the Hamiltonian is determined by the following list of parameters
$R,a,V,R_{SO},a_{SO},\tilde{V}$ and the reduced mass $\mu$; these
seven parameters \textit{define} the Woods-Saxon potential.

The parameters of the potential change as one goes over the nuclear
chart. The dependence of the above 7 parameters on the number of protons
and neutrons in the core-nucleon problem defines the \textit{structure
of parameterization}. In the following, we will describe our choice
of structure parameterization, which we call the {}``Seminole''
parameterization.

In the conventional parameterization of the Woods-Saxon potential,
as well as in our work, the size of nuclear potential is calculated
as \begin{equation}
R=R_{C}=R_{0}A^{1/3}\,,\quad R_{SO}=R_{0,SO}A^{1/3},\label{radii}\end{equation}
 in terms of the parameters $R_{0}$ and $R_{0,SO}$, which are constant
over the nuclear chart.

The surface diffuseness for both the central and spin-orbit potential
is assumed to be constant and size independent \begin{equation}
a=a_{SO}=const.\label{diffuse}\end{equation}

While the isospin dependence of the Coulomb force is obvious, the
behavior of the effective nuclear potential on the isospin of the
nucleon ${\bf t}$ and the core ${\bf T}'$ has to be introduced phenomenologically.
We adopted the suggestion by Lane \cite{lane62} to introduce an isospin
dependence to the potential by the lowest order isospin invariant
term

\begin{equation}
V=V_{0}\left(1-\frac{4\kappa}{A}\,\langle{\bf t}\cdot{\bf T}'\rangle\right).\label{isosp}\end{equation}

We chose the {}``minus'' sign so that the parameter $\kappa$ will
be consistent with conventions. For the ground-state of a nucleus,
the isospin quantum number is $T=|T_{z}|=|N-Z|/2$, which together
with the relation ${\bf t}+{\bf T}'={\bf T}$ leads to \begin{equation}
-4\langle{\bf t}\cdot{\bf T}'\rangle=\left\{ \begin{array}{lr}
3 & N=Z\\
\pm(N-Z+1)+2 & N>Z\\
\pm(N-Z-1)+2 & N<Z\end{array}\right.\,,\label{eq:iso}\end{equation}
 where here and below we use upper sign for the proton and the lower
sign for the neutron. Traditionally, the isospin-dependence of the
Woods-Saxon potential had been parameterized by the expression. \begin{equation}
V=V_{0}\left(1\pm\kappa\frac{(N-Z)}{A}\right)\label{eq:kappa}\end{equation}
 For heavy nuclei with large neutron excess, the difference between
the two definitions is small. However, the definition of Eq.~\ref{eq:iso}
leads to significantly different predictions in lighter nuclides around
N=Z. We discuss the validity of this assumption for the description
of our data in section \ref{sec:fit}.

The structure of the {}``Seminole'' parameterization is given by
Eqs. (\ref{isosp}) and (\ref{eq:iso}) which define the dependence
of the potential depth on $Z$ and $N$ in an isospin conserving way.

The spin-orbit interaction strength is determined by \begin{equation}
\tilde{V}=\lambda V_{0},\label{eq:lambdanew}\end{equation}
 using a proportionality constant $\lambda$ which is a constant,
making $\tilde{V}$ a constant. This aspect of our parameterization
is different from the traditional approach, where the spin orbit potential
depth is modified with the same isospin dependent factor as the central
potential. Finally, the reduced mass is given as in Eq.~ (\ref{eq:redmass}),
where the mass of the core is parameterized as $(A-1)$u. The 6 constants
$V_{0},R_{0},R_{0,SO},a=a_{SO},\lambda,$ and $\kappa$ and are the
actual parameters of the global {}``Seminole'' Woods-Saxon fit.

We have also studied a possible variation of the potential radius
with isospin. From our investigation of nuclear charge radii, which
will be described in section \ref{sec:radii}, we concluded that a
modification of potential radii with respect to the standard structure
of parameterization in Eq. (\ref{radii}) would lead to wrong predictions
of experimental charge and neutron radii.

\subsection{Existing parameterizations}

\begin{table*}
 \begin{tabular}{l||l|l|l|l|l|l|l}
Parameterization &
&
$V_{0}[MeV]$ &
$\kappa$ &
$R_{0}$ {[}fm] &
$a$ {[}fm]&
$\lambda$ &
$R_{0SO}${[}fm] \tabularnewline
\hline 
'Rost'\cite{Ros68} &
n &
49.6 &
0.86 &
1.347 &
0.7 &
31.5 &
1.28 \tabularnewline
&
p &
&
&
1.275 &
&
17.8 &
0.932 \tabularnewline
'Optimized'\cite{Dud79} &
n &
49.6 &
0.86 &
1.347 &
0.7 &
36 &
1.30 \tabularnewline
&
p &
&
&
1.275 &
&
36 &
1.30 \tabularnewline
'Universal'\cite{Dud82} &
n &
49.6 &
0.86 &
1.347 &
0.7 &
35 &
1.31 \tabularnewline
&
p &
&
&
1.275 &
&
36 &
1.32 \tabularnewline
'Chepurnov'\cite{Che67} &
&
53.3 &
0.63 &
1.24 &
0.63&
23.8 $^{*}$ &
1.24 \tabularnewline
'Wahlborn'\cite{Blo60} &
&
51 &
0.67 &
1.27 &
0.67&
32 &
1.27 \tabularnewline
\end{tabular}

\caption{Commonly used Woods-Saxon parameter sets in the literature. $^{*}$
The {}``Chepurnov'' parameterization introduced an additional isospin-dependence
for the spin-orbit interaction in deviation from the commonly used
proportionality to the central potential.}

\label{tab:litpars} 
\end{table*}

A number of parameterizations and parameter sets are available in the
literature. We list the most commonly used ones in Table
\ref{tab:litpars}.  All of these parameterizations use
Eq.~(\ref{eq:kappa}) for the isospin dependence of the nuclear
potential. In contrast to the {}``Seminole'' parameterization, the
same isospin dependence enters into spin-orbit term via
\begin{equation} \tilde{V}=\lambda V,\end{equation} which should be
compared with Eq.~(\ref{eq:lambdanew}).

The {}``Rost'' parameters \cite{Ros68} were determined from the
orbital energies of $^{208}$Pb. The {}``Optimized'' \cite{Dud79}
parameter set took the central potential parameters from ``Rost'', but
changed the spin-orbit interaction in order to improve predictions
for high-spin spectra in the lead region. A further refinement of
these parameters was introduced as the {}``Universal'' parameter
set, which improved the description of high spin states in $^{146}$Gd
\cite{Dud82}.

Common to these three parameterizations is the assumption of different
parameters sets for the proton and neutron spectra, and specifically
a larger neutron than proton potential radius. The implementation
of these parameterizations through the computer code \textsc{swbeta}
\cite{Cwi87} uses the the bare nucleon mass in the kinetic energy
operator and a reduced mass $\mu=\frac{A-1}{A}$ u in the spin-orbit
coupling potential.

\section{\label{sec:fit}Optimization of Woods-Saxon Potential Parameters}

\subsection{Calculation of the Woods-Saxon spectrum and adjustment of parameters}

Special attention had to be paid to the way in which the experimental
particle and hole states are compared to the Woods-Saxon calculation.
Single neutron states are naturally calculated as the potential for
$(N_m+1,Z_m)$, and single proton states as the $(N_m,Z_m+1)$, where
$N_m$ and $Z_m$ are the neutron and proton numbers of the respective
magic core. The hole states around the same magic nucleus are 
nucleons moving in the field of a core, which has one nucleon
less than the magic core, so that the orbitals based on hole states
are calculated at $(N_m,Z_m)$.

As a primary computational tool, we used a Woods-Saxon code which is a
part of the Continuum Shell Model code CoSMo discussed in \cite{CoSMo}
and that has been successfully used in studies of realistic nuclei
\cite{volya}. Web implementation of this code for bound states along
with the Seminole parameterization obtained as a result of this work
can be found in \cite{www}. The spectrum, radii, structure of
wave-functions and reaction calculations were performed using this
code which makes the current Woods-Saxon parameterization readily
available for further incorporation into the Continuum Shell Model.

The independent use of the publicly available program \textsc{swbeta}
\cite{Cwi87} to calculate the spectrum of the Woods-Saxon Hamiltonian
provided a crucial verification. This program was modified slightly
to incorporate the changes to the Woods-Saxon parameterization we
introduced. Note that, while \textsc{swbeta} calculates the spectrum
of a deformed Woods-Saxon potential, zero deformation was used for
all calculations presented here.

We also developed a parameter fit program, which calls the Woods-Saxon
calculation through a shell command with the hypothetical fit parameters
and reads back the calculated orbital spectrum. The actual parameter
optimization is realized through a Levenberg-Marquardt algorithm \cite{Lev44},
which minimizes the $\chi^{2}$ between the experimental and calculated
orbital energies by systematically varying the potential parameters.

While the spectrum of all nuclides listed in section \ref{sec:SPdata}
was usually fit with the same set of parameters, the fit program also
can allow for certain parameters to vary for each individual nuclide.
This option was used to systematically investigate how to improve
the dependence of certain parameters on the nuclide mass or the neutron-proton
asymmetry.

Armed with a fitting procedure we first consider a structure of the
{}``Universal'' parameterization and fit its nine parameters to
our data set of orbital energies around $^{16}$O, $^{40}$Ca, $^{48}$Ca,
$^{56}$Ni, $^{100}$Sn, $^{132}$Sn and $^{208}$Pb. The resulting
parameters called ``Fit A'' are listed in table \ref{tab:pars}
together with results from the {}``Universal'' parameters. We observe
that the large difference in central potential radius between neutron
and proton radii disappears in the fit, although these parameters
were free to adjust independently. The spin-orbit strength parameter
$\lambda$ is the only parameter in ``Fit A'' displaying substantially
asymmetric values for neutrons and protons. Table \ref{tab:pars}
also lists the RMS energy deviation obtained for the different nuclides.
The {}``Fit A'' parameters show a more consistent description of
the lighter nuclei than the {}``Universal'' parameters and a better
description of the nuclides with N=Z. The features of {}``Fit A'',
namely a symmetric parameterization for protons and neutrons, serve
as a guide to find an improved parameterization and parameter set
addressed below.  ``Fit B'', which is included in the
table will be discussed in the following section.

\begin{table*}[htbp]
 \begin{tabular}{l||l|l|l|l|l|l|l||l|l|l|l|l|l|l}
\multicolumn{8}{c||}{}&
\multicolumn{7}{c}{RMS energy deviation [MeV]}\tabularnewline
&
&
$V_{0}${[}MeV] &
$\kappa$ &
$R_{0}$ {[}fm] &
$a$ {[}fm] &
$\lambda$ &
$R_{0SO}${[}fm] &
\multicolumn{1}{c}{$^{16}$O}&
\multicolumn{1}{c}{$^{40}$Ca}&
\multicolumn{1}{c}{$^{48}$Ca}&
\multicolumn{1}{c}{$^{56}$Ni}&
\multicolumn{1}{c}{$^{100}$Sn}&
\multicolumn{1}{c}{$^{132}$Sn}&
\multicolumn{1}{c}{$^{208}$Pb}\tabularnewline
\hline 
Univ. &
n &
49.6 &
0.86 &
1.347 &
0.7 &
35 &
1.31&
2.65&
1.68&
1.07&
1.18&
1.20&
0.40&
0.37\tabularnewline
&
p &
&
&
1.275 &
&
36 &
1.32&
2.31&
2.18&
1.63&
1.40&
2.29&
0.61&
0.43\tabularnewline
Fit A &
n &
52.78&
0.636 &
1.282 &
0.62&
24.0&
1.18&
1.91&
0.84&
1.06&
0.55&
0.72&
0.53&
0.60\tabularnewline
&
p &
&
&
1.288 &
&
21.2&
1.05&
1.96&
0.69&
0.64&
0.44&
1.31&
0.43&
0.38\tabularnewline
Fit B &
&
var &
0.000 &
1.260 &
0.68&
25.5&
1.21&
0.33&
0.52&
1.34&
0.45&
0.74&
0.53&
0.40\tabularnewline
&
&
var &
&
&
&
&
&
0.32&
0.58&
0.76&
0.39&
1.19&
0.34&
0.32\tabularnewline
\end{tabular}

\caption{Parameters and fit quality of the {}``Universal'' parameterization
compared to parameter fit {}``Fit A'' adjusting the same parameters
to the orbital energy data presented in section \protect\ref{sec:SPdata}.
In {}``Fit B'' parameters labeled {}``var'' were allowed individual
values for each nuclide for protons, neutrons, for particle and hole
state (see text). The individual parameter values are displayed in
Fig. \protect\ref{fig:vnull_iso}. The values of the other parameters
were allowed to adjust one common value during the fit, with the exception
of $\kappa$, which was held constant at 0.}

\label{tab:pars} 
\end{table*}

\subsection{Changes to the parameterization}

\begin{figure}[htbp]
\includegraphics[width=6cm,angle=-90]{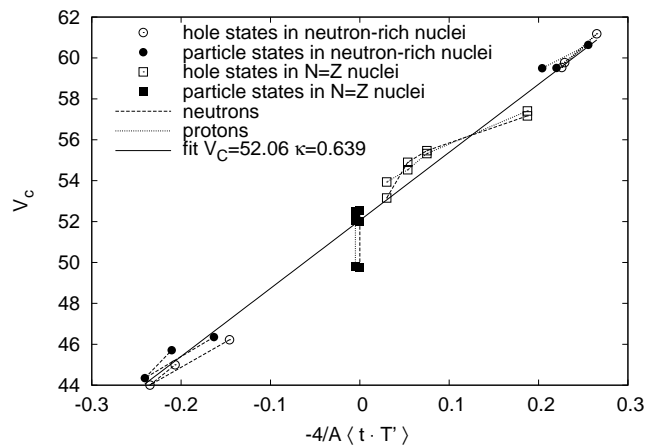}

\caption{Central potential strength extracted from {}``Fit B'' as a
function of the isospin factor $-\frac{4}{A}\,\langle{\bf t}\cdot{\bf
T}'\rangle$ (see Eq.~\ref{eq:iso}). The straight line represents the
central potential parameterization assumed in this work. }

\label{fig:vnull_iso} 
\end{figure}

In this paragraph we investigate the validity of the isospin-conserving
parameterization introduced in Eq.~(\ref{eq:iso}). In the procedure
listed as {}``Fit B'' in table \ref{tab:pars}, we introduced the
reduced mass of Eq.~(\ref{eq:redmass}), and the constant spin-orbit
potential strength of equation (\ref{eq:lambdanew}). The central
potential depth was optimized with a separate value for each neutron
and proton, particle and hole data set in each nuclide, a total of
28 values. The other parameters $R_{0}$, $R_{0,SO}$ $a$ and $a_{SO}$
were free to adjust one common value for all nuclides. The extracted
$V$ values are displayed in Fig.\ref{fig:vnull_iso}, plotted as
a function of the isospin multiplier $-\frac{4}{A}\,\langle{\bf t}\cdot{\bf T}'\rangle$.
The values show a linear dependence on the isospin factor, in agreement
with the parameterization (\ref{eq:iso}). The only significant deviations
from the linear dependence are found for both the particle ($\frac{4}{A}\,\langle{\bf t}\cdot{\bf T}'\rangle$
= 0,$V_{0}=49.8$ MeV ) and hole orbitals ($\frac{4}{A}\,\langle{\bf t}\cdot{\bf T}'\rangle$
= 0.1875, $V=57.3$ MeV) of Oxygen. These data points are an indication
that for lighter nuclei a reduced potential depth would improve the
fit quality. However, we did not introduce a mass dependence to the
potential depth parameter for the sake of simplicity and a more stable
extrapolation.

The question, whether the spin-orbit potential depends on isospin was
investigated by allowing a modification in analogy to the
isospin-dependence of the central potential. Note that in the previous
parameterizations \cite{Dud82}, the spin orbit potential strength was
assumed to be directly proportional to the central potential, thus
having an explicit isospin-dependence. As was stated earlier the
isospin-dependence of a spin-orbit term, although an interesting
question, can not be fully addressed with our approach.  All our
attempts to adjust this parameter led to a slight deterioration of the
fit with most effects due to the $({\bf T}\cdot {\bf t})({\bf
l}\cdot{\bf s})$ term being subsumed in the adjustment of the other
parameters. Therefore, we defined our parameterization as not
containing this term. We believe that this feature given its marginal
and elusive effect on the single particle bound and resonant spectra
is not essential for our purpose. In fact the isospin dependence in
the spin orbit is a weak manifestation of the physics that goes beyond
the mean field and thus it should be treated using a more advanced
approach such as shell model based on the Woods-Saxon basis from our
parameterization. Furthermore, the reduced number of free parameters
allows us to achieve a more stable parameterization.

\subsection{Charge Radii}

\label{sec:radii} While the description of the orbital energy spectra
is an important component of any model of single particle motion,
the Woods Saxon potential also strives to reproduce the geometry of
the nucleus and properties that are sensitive to the geometry of its
wave functions. Especially since the radial and potential depth values
parameters show strong correlations in their influence on the particle
energy spectrum, it is important to study the geometric properties
of the wave-functions in an independent fashion.

We took data on experimental rms charge radii from the compilation
by I. Angeli \cite{Ang04}, for the proton-magic nuclides $^{16}$O
- $^{18}$O, $^{40}$Ca - $^{50}$Ca, $^{108}$Sn - $^{115}$Sn and
$^{190}$Pb - $^{213}$Pb. Using our new parameterization, we calculated
the RMS radii of the same nuclides by averaging in square all proton
orbital radii with their proper occupation numbers up to the Fermi
level. To the orbital RMS radius, the RMS radius of the proton was
added in square by $R_{charge}=\sqrt{R_{orbit}^{2}+R_{proton}^{2}}$.

\begin{figure*}[htbp]
\includegraphics[width=12cm,angle=-90]{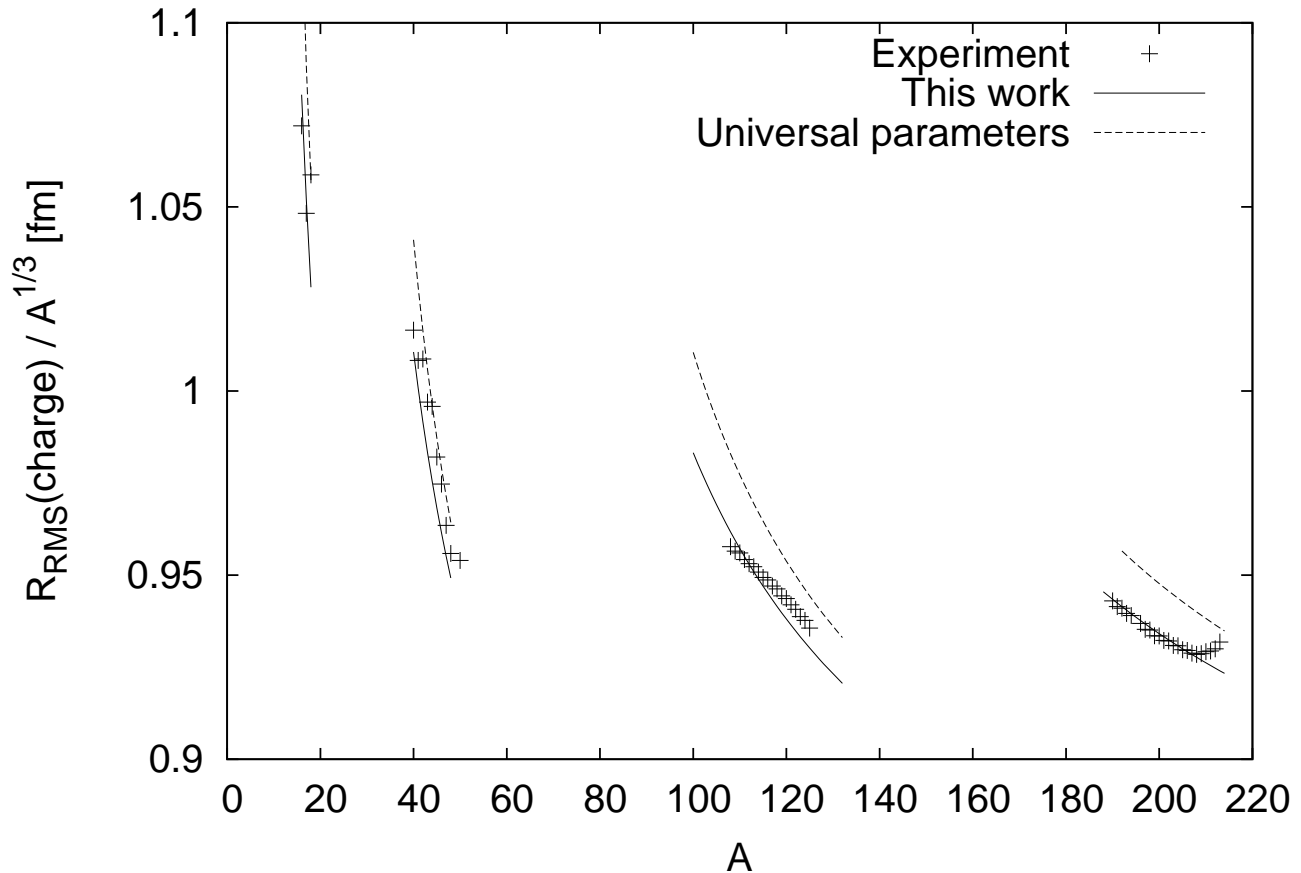}

\caption{Experimental rms charge radii divided by the scaling factor A$^{1/3}$
as a function of mass number (symbols), for a set of Oxygen, Calcium,
Tin and Lead isotopes. The data is compared to the results calculated
with our parameters (solid line) and the {}``Universal'' parameters
(dashed line). The experimental uncertainties are typically smaller
than 0.005 fm.}

\label{fig:radii} 
\end{figure*}

The experimental charge radii, the results obtained for our parameters
and the results for the {}``Universal'' parameters are displayed
in Fig.~\ref{fig:radii}. Note that for each isotopic chain displayed
here, the protons occupy the same orbitals, so that the observed variations
are due to changes in the orbital geometry only.

We varied the potential radius parameter for values between 1.25 and
1.29 fm, while adjusting the other potential parameters for each step
to reproduce the orbital energy data. The value of $R_{0}=1.26$ fm
best reproduces the charge radii for the $^{190}$Pb-$^{208}$Pb
isotopes. The increase of experimental charge radii observed beyond
$^{208}$Pb has been discussed in the context of various models, most
recently in a systematic discussion of quadrupole correlation effects
in a self-consistent calculation with the SLy4 interaction
\cite{Ben06}. Our much simpler model can not reproduce this increase,
either. The Tin and Calcium isotopes exhibit slightly larger rms radii
than calculated. It is likely that this deviation is due to an onset
of deformation in mid-shell, consistent with the calculations of
\cite{Ben06}.

It should be noted, that the variation of rms radii within the isotopic
chains is determined only in part by the global scaling of the potential
radius with $A^{1/3}$. Clearly evident in the data is a systematic
dependence on the binding energy of the fermi orbital, where proton-rich
nuclides have relatively larger charge radii than expected from the
global scaling of the potential radius. Our calculations illustrate,
that the same orbitals occupy a larger fraction of the potential radius
as they become less bound.

\begin{figure}[htbp]
\includegraphics[width=8cm]{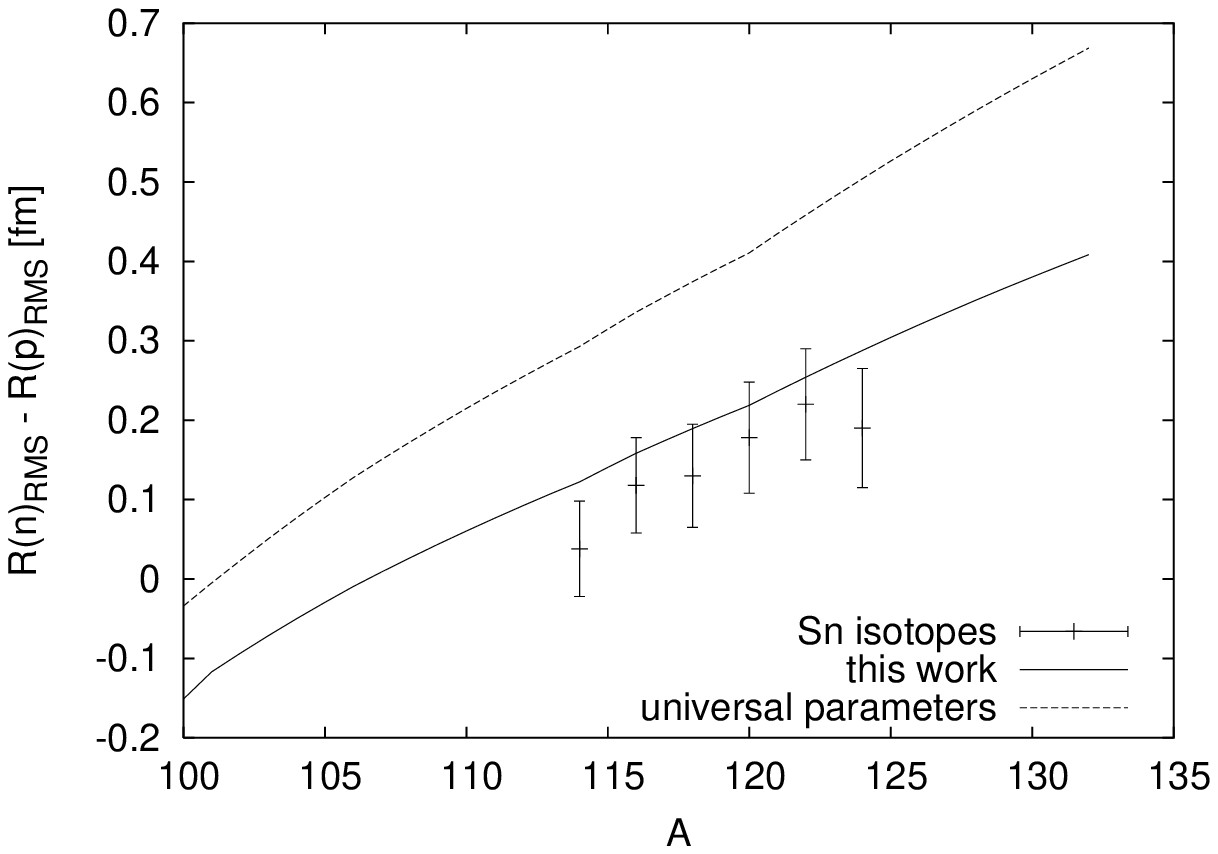}

\caption{Values for the difference between the neutron and the proton rms
radii in Sin isotopes, extracted from experimental data in spin-dipole
resonances \protect\cite{Kra99} (symbols). The data is compared
to the results calculated with our fit parameters (solid line) and
the {}``Universal'' parameters (dashed line).}

\label{fig:radii_pndiff} 
\end{figure}

Direct data on the neutron radii is much harder to obtain; However,
some information on the relative radii of neutrons and protons can
be deduced from isovector spin-dipole resonances, studied e.g. with
the $(^{3}$He$,t)$ reaction at intermediate energies. In Ref. \cite{Kra99},
the authors extracted values for the difference between neutron and
proton rms radii for even-mass Tin nuclides, which we display in Fig.~\ref{fig:radii_pndiff},
along with the values calculated for our Woods-Saxon parameterization
and the {}``Universal'' parameterization. The data presented in
the figure supports the assumption of symmetric proton and neutron
potential radii rather than the relatively larger neutron potential
of the {}``Universal'' parameterization.

\section{\label{sec:Results} The new Woods-Saxon Potential Parameterization}

\subsection{The Parameters}

We arrive at the structure parameterization given by Eqs.~(\ref{eq:redmass}),
(\ref{radii})-(\ref{eq:lambdanew}). The corresponding parameters
were adjusted to the orbital energy data. Their values are displayed
in table \ref{tab:newpars}

\begin{table}[htbp]
 \begin{tabular}{c|c|c|c|c|c|c}
$V_{0}$ {[}MeV]&
$\kappa$ &
$R_{0}$ {[}fm]&
$a=a_{SO}$ {[}fm] &
$\lambda$ &
$R_{0,SO}$ {[}fm]&
\tabularnewline
\hline 
52.06 &
0.639 &
1.260 &
0.662 &
24.1 &
1.16&
\tabularnewline
\end{tabular}

\caption{Woods-Saxon potential parameter set obtained in this work. }

\label{tab:newpars} 
\end{table}

Summarizing, the important properties of the parameterization and
the parameter set are: 

\begin{itemize}
\item Reduced mass in the form given by Eq.~(\ref{eq:redmass}) assures
correct two-body kinematics of nucleon and core. The average mass
per nucleon in a nucleus is assumed to be equal to an atomic mass
unit $u=931.49$ MeV/c$^{2}$. The introduction of the reduced mass
led to a substantial improvement in the quality of description for
the Oxygen and Calcium spectra. 
\item Identical potential parameters for neutrons and protons, which in
conjunction with Eq.~(\ref{isosp}) assures isospin conservation. 
\item The spin-orbit term is constant and does not depend on isospin. 
\item There are only six parameters that were fitted. These parameters are
given in table \ref{tab:newpars}. 
\end{itemize}

\subsection{\label{sec:Discussion}Discussion of Fit Results}

\begin{figure*}[p]
\includegraphics[width=14cm]{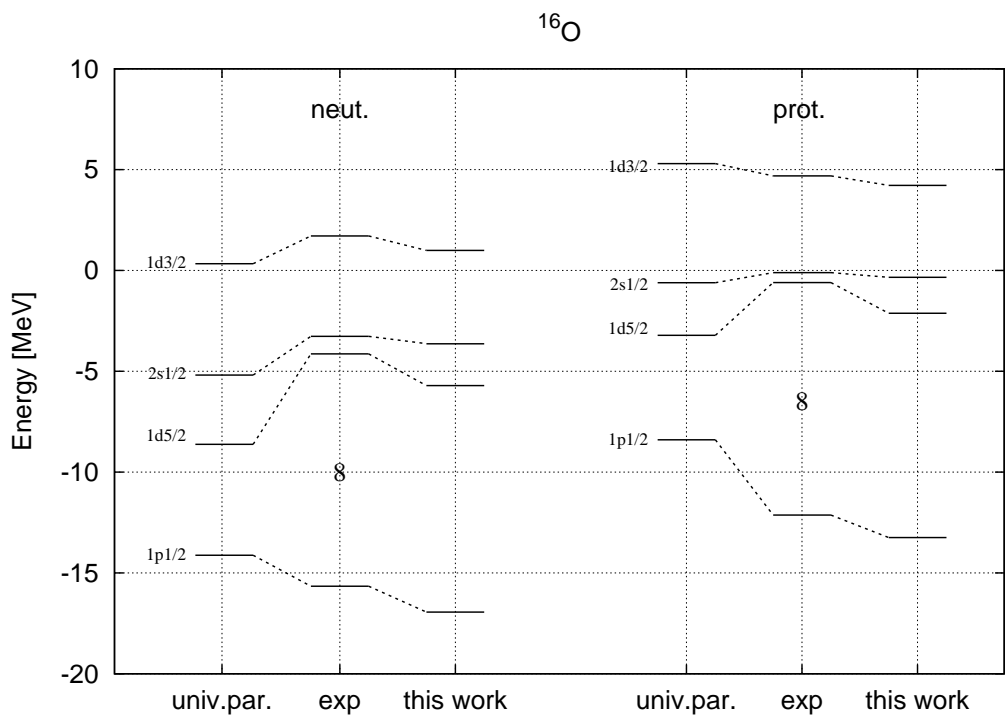}

\caption{Orbital energy spectrum for $^{16}$O. }

\label{fig:O16} 
\end{figure*}

\begin{figure*}[p]
\includegraphics[width=14cm]{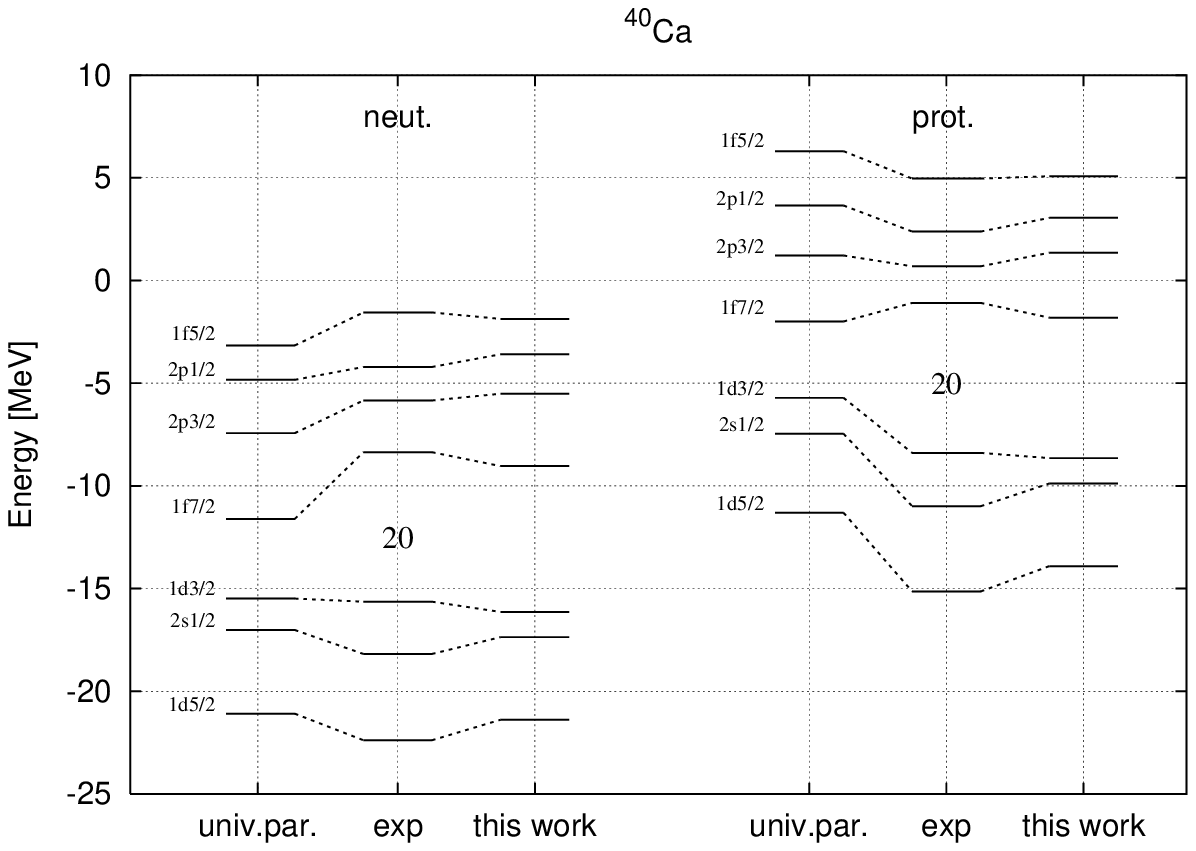}

\caption{Orbital energy spectrum for $^{40}$Ca.}

\label{fig:Ca40} 
\end{figure*}

\begin{figure*}[p]
\includegraphics[width=14cm]{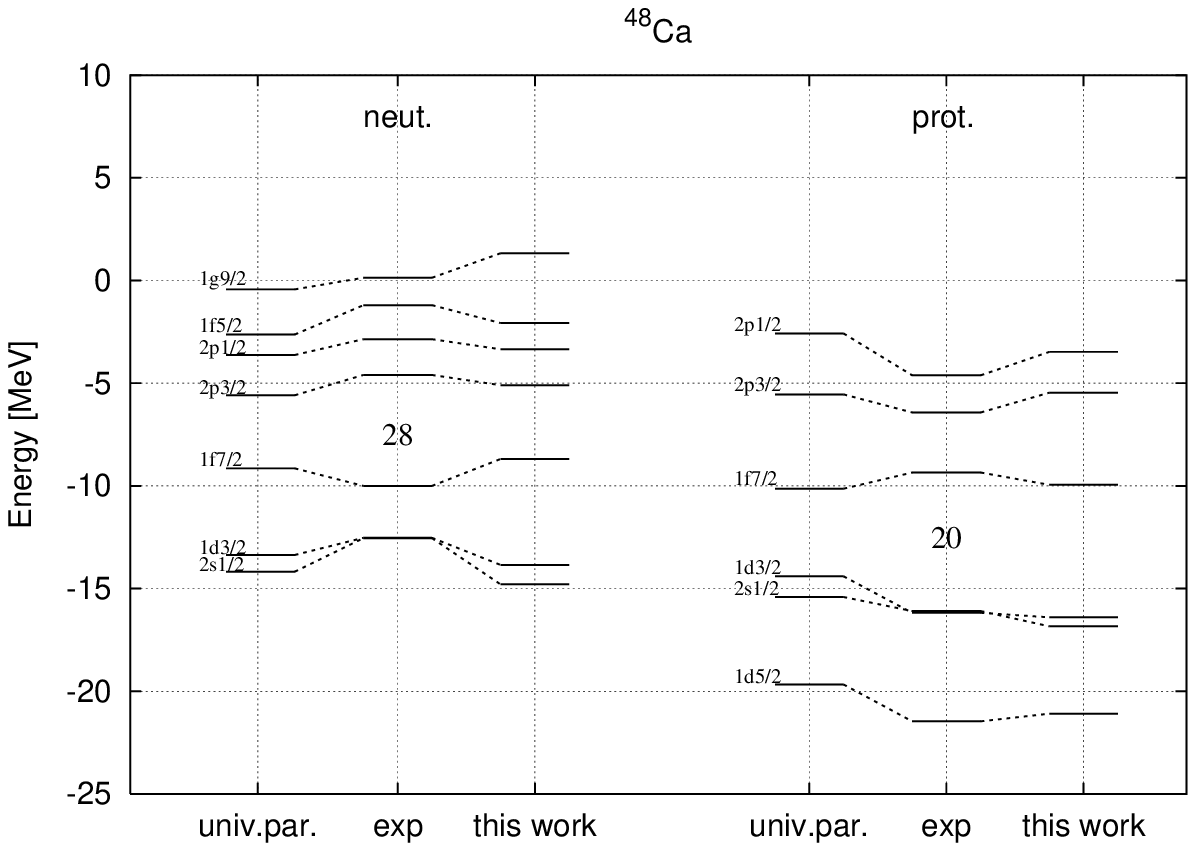}

\caption{Orbital energy spectrum for $^{48}$Ca.}

\label{fig:Ca48} 
\end{figure*}

\begin{figure*}[p]
\includegraphics[width=14cm]{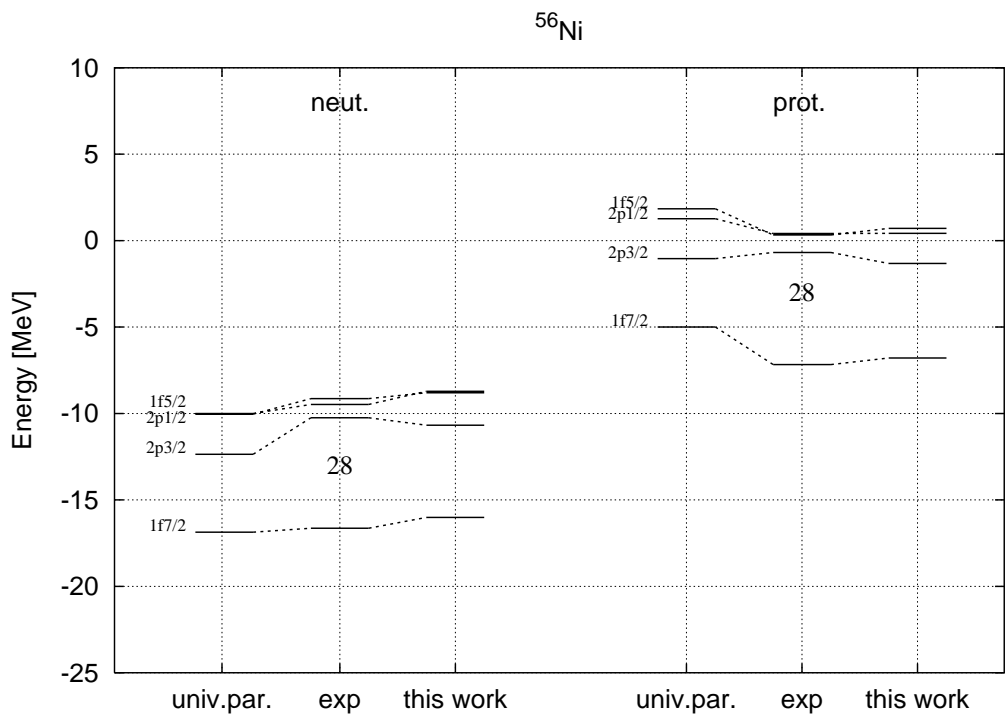}

\caption{Orbital energy spectrum for $^{56}$Ni.}

\label{fig:Ni56} 
\end{figure*}

\begin{figure*}[p]
\includegraphics[width=14cm]{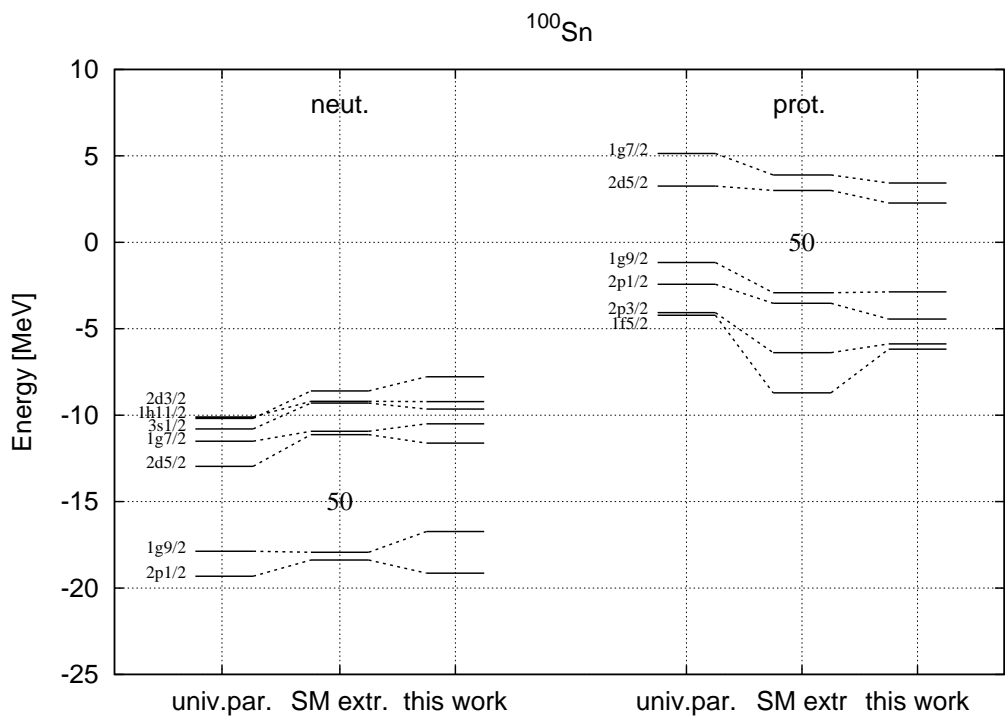}

\caption{Orbital energy spectrum for $^{100}$Sn. The data in the
center column is extracted from systematic Shell-model calculations of
nuclides approaching $^{100}$Sn\protect\cite{Gra95}. }

\label{fig:Sn100} 
\end{figure*}

\begin{figure*}[p]
\includegraphics[width=14cm]{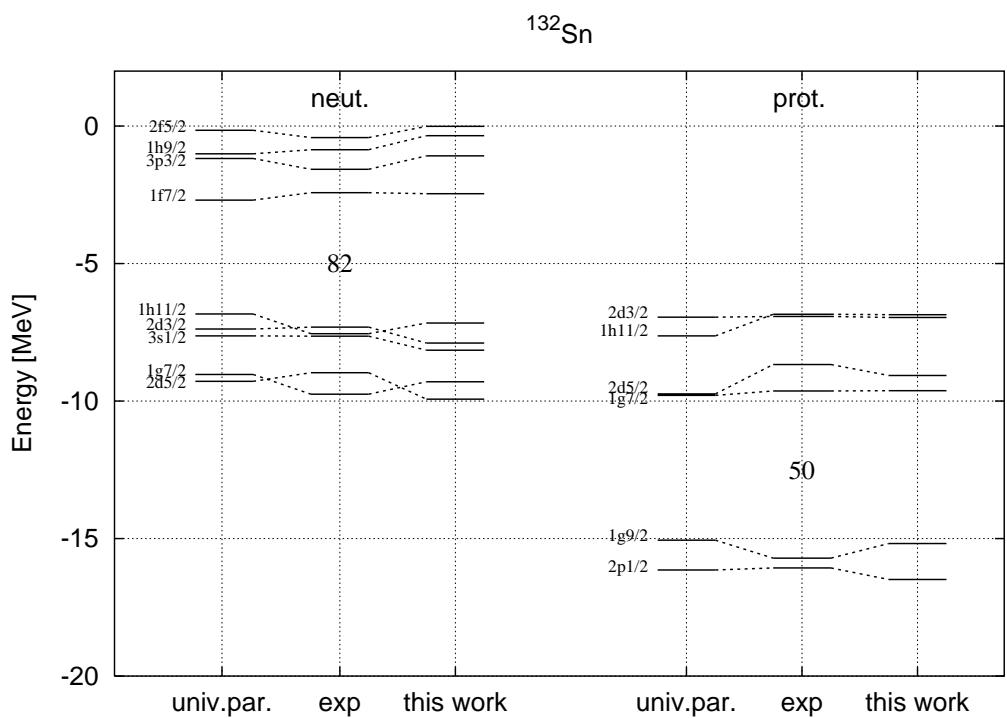}

\caption{Orbital energy spectrum for $^{132}$Sn.}

\label{fig:Sn132} 
\end{figure*}

\begin{figure*}[p]
\includegraphics[width=14cm]{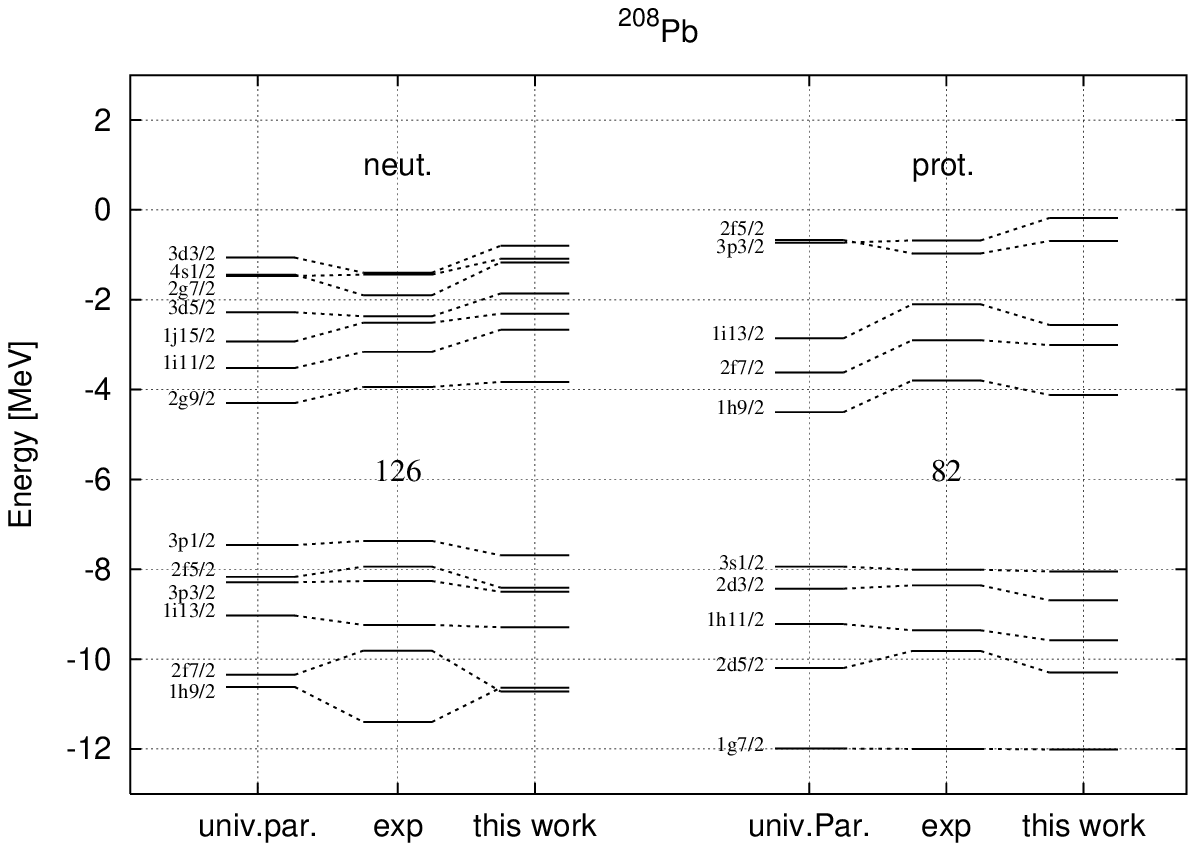}

\caption{Orbital energy spectrum for $^{208}$Pb.}

\label{fig:Pb208} 
\end{figure*}

The experimental and calculated orbital energy spectra are displayed
in figures \ref{fig:O16}-\ref{fig:Pb208}. We will first address
the spectra of the symmetric N=Z nuclides $^{16}$O, $^{40}$Ca, $^{56}$Ni
and $^{100}$Sn. For these nuclides, the spectrum calculated with
the {}``Universal'' parameterization shows consistently too small
neutron and proton energy gaps around the magic numbers 8,20,28 and
50. The new parameterization greatly improves the accuracy of the
orbital spectrum description.

As mentioned in section \ref{sec:WSCalc}, the {}``Universal'' parameterization
was primarily developed from the spectrum of $^{208}$Pb. It is therefore
not surprising that it is less applicable to symmetric N=Z nuclides,
especially since it is based on substantially different neutron and
proton potential parameters. The {}``Universal'' parameters fare
better with the the neutron-rich nuclides $^{48}$Ca, $^{132}$Sn
and $^{208}$Pb, which all have similar neutron-excess values of $(N-Z)/A=0.167,0.242$
and $0.212$, respectively. In general, the description of these nuclides
by the new parameters is of similar quality as the {}``Universal''
parameterization, with one exception -- although the RMS deviation
of the neutron particle spectrum of $^{208}$Pb with the new parameters
(0.5 MeV) is similar to the one obtained from the {}``Universal''
parameters (0.37 MeV), the new parameter spectrum calculates all neutron
{}``particle'' orbitals systematically less bound than the experimental
values. The magnitude of these deviations is not beyond the typical
discrepancies of individual orbital fits in other nuclides, so that
we do not think it represents a systematic problem within our parameterization.
It should also be noted, that we could not find experimental information
on the fragmentation of the single-particle strength for the nuclide
in question, $^{209}$Pb, in the literature. If we assume that the
orbital fragmentation of high-lying states in $^{209}$Pb is similar
to that of the neutron hole orbitals in $^{207}$Pb, the systematic
{}``under-binding'' of our description would vanish.

\section{Predictions of Shell Structure in Exotic Nuclides}

The discussion of all consequences of our new parameterization for the
Woods-Saxon potential is clearly beyond the scope of this paper.  The
reader may further explore the features of this parameterization using
our web-based program \cite{www}.  Here, we want to briefly discuss
the general properties of the shell structure for exotic nuclides and
the limits of nuclear binding it implies. We use the {}``Seminole''
parameterization to calculate shell gap energies by filling the
orbitals to the Fermi level and calculating the ernergy gap to the
next unoccupied orbital for all nuclides between Z=8 and Z=150. Table
\ref{tab:magic} lists the nuclides, for which a shell gap of at least
1.5 MeV is present for both neutrons and protons and for which both
protons and neutrons were calculated to be bound. Note, that in this
table, the gap is calculated from the orbital energy differences within
one potential calculation, while the data represented in each of the
figures \ref{fig:O16}--\ref{fig:Pb208} corresponds to two
calculations, one for the core and one for the odd-nucleon neighbor,
which leads to slightly different shell-gap values. 
Table \ref{tab:magic} also lists the neutron and proton
separation energy, calculated as the binding energy of the Fermi level.
Experimental values are taken from the compilation of \cite{Aud03}.

\begin{figure*}[htbp]
\includegraphics[width=1\textwidth]{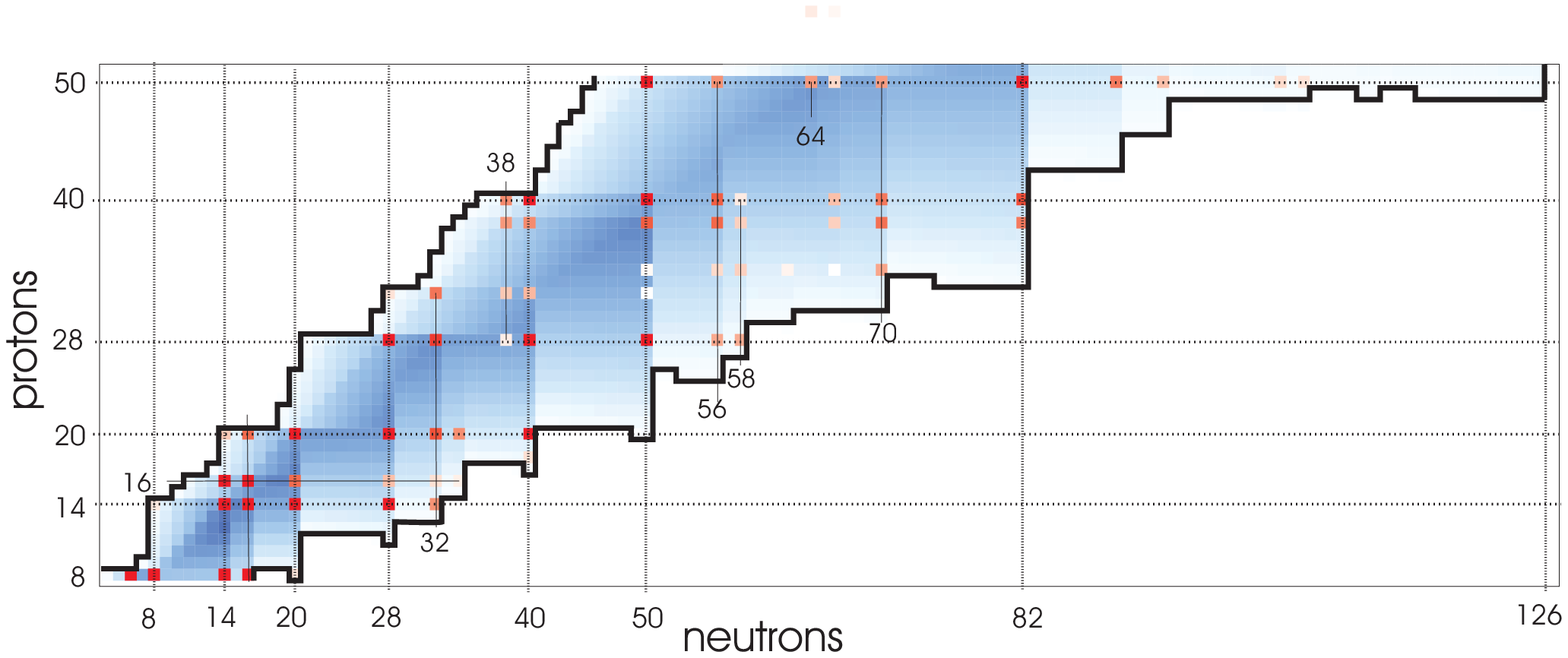}

\caption{Nuclear chart representing the calculated shell gaps, nucleon separation
energies and drip-lines of nuclides between nuclides of Z=8 and Z=51.
For doubly sub-magic nuclei, which are located at the crossing points
of the dotted lines, the minimum of the proton and neutron shell gaps
is represented by the shade of red, where larger gaps correspond to
a deeper color. All other bound nuclides are represented by a shade
of cyan color, which represents the minimum of the proton and neutron
separation energies. The calculated proton and neutron drip-lines
are also drawn. }

\label{fig:nucchart} 
\end{figure*}

The data for nuclides between Z=8 and Z=51 is also represented as
a nuclear chart in Figure \ref{fig:nucchart}. For nuclides with doubly
closed sub-shells, the minimum of the respective proton and neutron
shell gaps is represented by shades of red. The nucleon numbers of
sub-shell closures are indicated by thin lines. All other nuclides
are represented by shades of cyan color, which indicates the minimum
of separation energy in protons and neutrons. We also draw the calculated
limits of nuclear binding.

We find it necessary to stress again that Woods-Saxon parameterization
is a very crude approximation to the nuclear mean field; The true
potential can be quite different (the possibility of deformation is
particulary noteworthy) and could include more complicated spin and
tensor structures. Furthermore the parameters of potential can change
as a function of $N$ and $Z$ in a non-uniform manner \cite{sak}.
Although our potential is intended for use as a part of a more involved
many-body approach it is still instructive to explore the shell evolution
due to mere geometric modification that appear in our parameterization.

%
\begin{table}[htbp]
 \begin{tabular}{l|l|l|r|r|r|r|r|r}
&
&
&
WS &
WS &
WS &
WS &
exp &
exp \tabularnewline
&
N&
Z&
Gap$_{N}$&
Gap$_{P}$&
S$_{N}$&
S$_{P}$&
S$_{N}$&
S$_{P}$\tabularnewline
&
&
&
{[}MeV]&
{[}MeV]&
{[}MeV]&
{[}MeV]&
{[}MeV]&
{[}MeV]\tabularnewline
\hline 
$^{14}$O &
6 &
8 &
5.64 &
7.01{*}&
23.80 &
5.57 &
23.18 &
4.63 \tabularnewline
$^{16}$O &
8 &
8 &
8.97 &
8.56 &
16.93 &
13.24 &
15.66 &
12.13\tabularnewline
$^{22}$O &
14 &
8 &
2.23 &
9.58 &
5.58 &
24.78 &
6.85 &
23.26\tabularnewline
$^{24}$O &
16 &
8 &
3.51{*}&
9.68 &
3.29 &
27.78 &
3.6 &
26.6{*}{*}\tabularnewline
$^{28}$O &
20 &
8 &
3.82{*}&
9.75 &
0.23 &
32.88 &
&
\tabularnewline
$^{28}$Si &
14 &
14 &
3.81 &
3.71 &
16.22 &
10.67 &
17.18 &
11.58\tabularnewline
$^{30}$Si &
16 &
14 &
2.39 &
3.95 &
10.58 &
13.27 &
10.61 &
13.51\tabularnewline
$^{34}$Si &
20 &
14 &
5.12 &
4.24 &
7.94 &
17.94 &
7.53 &
18.72\tabularnewline
$^{42}$Si &
28 &
14 &
2.23 &
4.44 &
3.20 &
25.50 &
3.20{*}{*}&
24.56{*}{*}\tabularnewline
$^{30}$S &
14 &
16 &
3.97 &
2.46 &
18.78 &
4.51 &
18.98 &
4.40\tabularnewline
$^{32}$S &
16 &
16 &
2.04 &
2.02 &
14.23 &
8.07 &
15.04 &
8.86\tabularnewline
$^{40}$Ca &
20 &
20 &
5.95 &
5.43 &
16.15 &
8.66 &
15.64 &
8.33\tabularnewline
$^{48}$Ca &
28 &
20 &
3.53 &
5.95 &
8.71 &
16.39 &
9.94 &
15.86\tabularnewline
$^{52}$Ca &
32 &
20 &
1.67 &
6.10 &
5.02 &
19.67 &
4.72 &
17.80{*}{*}\tabularnewline
$^{60}$Ca &
40 &
20 &
2.96{*}&
5.95 &
2.53 &
24.96 &
&
\tabularnewline
$^{70}$Ca &
50 &
20 &
2.22{*}&
5.63 &
0.25 &
30.25 &
&
\tabularnewline
$^{56}$Ni &
28 &
28 &
4.44 &
4.35 &
16.01 &
6.79 &
16.64 &
7.17\tabularnewline
$^{68}$Ni &
40 &
28 &
3.29 &
4.97 &
8.16 &
14.86 &
7.79 &
15.69\tabularnewline
$^{78}$Ni &
50 &
28 &
3.55 &
5.13 &
5.04 &
20.30 &
5.62{*}{*}&
\tabularnewline
$^{88}$Sr &
50 &
38 &
4.53 &
1.57 &
10.50 &
10.62 &
11.12 &
10.61\tabularnewline
$^{80}$Zr &
40 &
40 &
2.92 &
2.29 &
15.49 &
3.20 &
16.23 &
4.44\tabularnewline
$^{90}$Zr &
50 &
40 &
4.66 &
2.06 &
11.53 &
8.19 &
11.97 &
8.35\tabularnewline
$^{96}$Zr &
56 &
40 &
1.57 &
1.96 &
6.68 &
10.91 &
7.86 &
11.52\tabularnewline
$^{122}$Zr &
82 &
40 &
3.98{*}&
1.69 &
3.38 &
20.60 &
&
\tabularnewline
$^{100}$Sn &
50 &
50 &
5.15 &
5.61{*}&
16.74 &
2.87 &
17.65{*}{*}&
2.80{*}\tabularnewline
$^{132}$Sn &
82 &
50 &
4.75 &
5.23 &
7.19 &
15.16 &
7.31 &
15.71\tabularnewline
$^{146}$Gd &
82 &
64 &
5.27 &
1.69 &
12.04&
4.72 &
11.2 &
5.38 \tabularnewline
$^{182}$Pb &
100&
82 &
1.53 &
4.43{*}&
11.04&
2.01 &
11.75 &
1.31 \tabularnewline
$^{208}$Pb &
126&
82 &
3.88 &
3.64 &
7.69 &
8.00 &
7.37 &
8.01 \tabularnewline
$^{218}$U &
126&
92 &
3.82 &
1.16 &
9.81 &
1.78 &
8.85 &
2.43 \tabularnewline
$^{256}$U &
164&
92 &
1.37 &
1.68 &
5.06 &
9.64 &
&
\tabularnewline
$^{276}$U &
184&
92 &
2.39 &
1.81 &
2.68 &
13.11 &
&
\tabularnewline
$^{278}$114&
164&
114 &
1.98 &
1.97 &
9.28 &
2.19 &
&
\tabularnewline
$^{298}$114&
184&
114 &
2.84 &
1.87 &
6.10 &
5.41 &
&
\tabularnewline
$^{342}$114&
228&
114 &
2.09 &
1.70 &
3.14 &
11.50 &
&
\tabularnewline
\end{tabular}

\caption{Table of nuclides with a Fermi level shell gap calculated larger
than 1.5 MeV in both protons and neutrons by the {}``Seminole''
parameterization. Cases where the upper orbital for the shell gap
was calculated to be unbound are marked with an asterisk. The following
columns list the calculated neutron and proton separation energies
and the experimental separation energies from \protect\cite{Aud03}.
Values marked with two asterisks are based on extrapolation, not experiment. More calculations 
may be performed
using our program at \cite{www}.}

\label{tab:magic} 
\end{table}

\subsection{N=16 and N=20 Shell Closures}

\label{sec:N20} The modification of shell structure in exotic nuclides
is currently a topic on which great experimental and theoretical efforts
are expended. Otsuka and co-workers have attributed a significant
renormalization of shell model single-particle energies to scalar
\cite{Ots01} and tensor \cite{otsuka05} components of the spin-spin
proton-neutron residual interaction. These interactions act mainly
on the spin-orbit partners of protons and neutrons. For instance,
as the proton $d_{5/2}$ occupation goes from 6 protons at $_{14}^{30}$Si$_{16}$
to zero for the $_{8}^{24}$O$_{16}$, the effective orbital energy
for the neutron $d_{3/2}$ orbital is calculated to increase. This
effect creates a N=16 shell gap at the neutron drip line and also
leads to the breaking of the N=20 magic number in the {}``island
of inversion'' around $^{32}$Mg.

Our model naturally includes halo phenomena of weakly bound states,
as observed in the charge radii displayed in Fig.~\ref{fig:radii}.
In consequence, the increased radial extension leads to modifications
in the spin-orbit splitting. The corresponding (sub)-shell gaps for
the above mentioned example are included in table \ref{tab:magic}
-- the calculated N=16 shell gap increases from 2.43 MeV in $^{30}$Si
to 4.51 MeV in $^{24}$O, while the N=20 gap decreases from 5.09 MeV
to 3.82 MeV between $^{34}$Si and $^{28}$O. 

We conclude that some modifications of shell structure in exotic
nuclides can be explained by a model as simple as the presented
Woods-Saxon potential, as it includes quantum-mechanical dynamics of
weakly bound particles. In our results however, the size of the N=16
shell gap does not reach the $\approx6$~MeV calculated in
\cite{Ots01}.

\subsection{N=28 and Z=14 Shell Closures}

The question, whether also the N=28 shell closure is broken in neutron-rich
nuclides is another topic of intense investigations and discussion.
Here, the mechanics of a possible breaking of the N=28 gap would be
different from the N=20 case discussed above -- since the N=28 closure
is based on the $f_{7/2}$ orbital and the proton $f_{5/2}$ remains
empty for the neutron-rich as well as the stable N=28 nuclides.

The Woods-Saxon potential calculation shows a clear decrease of the
N=28 gap as we compare the values for the increasingly exotic $^{56}$Ni,
$^{48}$Ca and $^{42}$Si, at 4.44 MeV, 3.53 MeV and 2.22 MeV, respectively
(see Tab.\ref{tab:magic}). Based on expectations of a decreasing
N=28 shell gap, several theoretical investigations \cite{Lal98}--\cite{Rod02}
had predicted a strongly deformed ground state in $_{14}^{42}$Si$_{28}$.
However, in an experiment of one- and two-proton knockout reactions
with a $^{44}$S beam, it was shown that there is a substantial Z=14
shell closure \cite{Fri05,Fri06} and nearly degenerate $s_{1/2}$
and $d_{3/2}$ orbitals. The shell gap data listed in table \ref{tab:magic}
and represented in Fig.~\ref{fig:nucchart} supports these findings;
The calculated Z=14 sub-shell gap increases from 3.68 MeV in the stable
$^{28}$Si to 4.42 MeV in $^{42}$Si. Our calculation also shows the
near degeneracy of the $s_{1/2}$ and $d_{3/2}$ orbitals, which here
are calculated only 0.63 MeV apart. Note that this orbital spacing
is also present in the $^{48}$Ca data displayed in figure \ref{fig:Ca48}.
This spacing generates a large $Z=14$ gap and a very small $Z=16$
gap, which is consistent with a magic character of $^{42}$Si with
a stabilized N=28 closure and a moderate deformation $^{44}$S, as
observed in \cite{Sch96,Gla97}.

\subsection{Limits of Nuclear Binding}

The limits of neutron binding and location of the neutron drip-line
is a very important property of atomic nuclei, which is largely unknown
at this time. The location of the neutron dripline has implications
for the nucleosynthesis of heavy elements in the universe as well
as for the physics of neutron stars. Radioactive beam facilities of
the next generation are designed to establish the limits of neutron
binding as one of their central goals.

We therefore want to discuss some of the straightforward implications
of our Woods-Saxon parameterization for the limits of nuclear binding.
It is clear that a model like the Woods Saxon-potential does not include
residual interactions and therefore a lot of interesting physics is
out of reach for this approach. Nevertheless it can serve as the first
order approximation in the determination of the drip lines and the
basis for more detailed theoretical investigations.

The calculated proton and neutron drip-lines are displayed in Fig.\ref{fig:nucchart}
and clearly show the effect of shell and sub-shell closures. Interestingly,
the neutron drip-line shows a number of instances, e.g. the Ca isotopes,
where the presence of magic numbers leads to regions, where an element
has bound nuclides after unbound ones. This {}``meandering'' of
the neutron drip line does not occur for the proton drip line.

Oxygen is the heaviest element, for which the location of the neutron
drip line is known at the present time. Experiments with high sensitivity
have established, that $^{26}$O is unbound \cite{Gui90,Fau95}. Our
calculation, represented in Fig.~\ref{fig:nucchart} also shows $^{26}$O
to be unbound, but brings back $^{28}$O as a nuclide bound by 0.21
MeV due to the N=20 shell closure.

\begin{figure}[htbp]
\includegraphics[width=8cm]{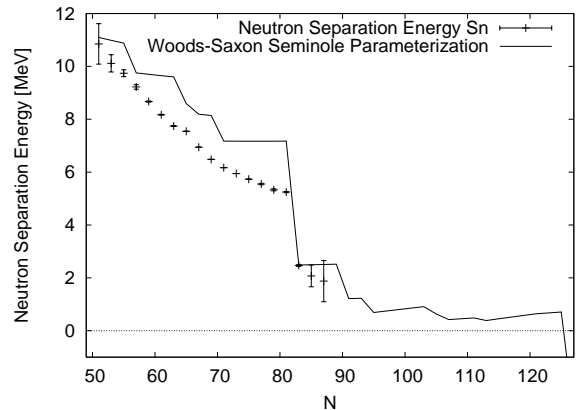}

\caption{Neutron separation energies for odd-N Tin isotopes, calculated with
the {}``Seminole'' parameterization of the Woods-Saxon potential
and experimental values from \protect\cite{Aud03}. }

\label{fig:bind_sn_n} 
\end{figure}

The Tin isotopes are calculated to be neutron-bound up to $^{176}$Sn.
We display the calculated and experimental neutron separation energies
for the odd-neutron Sn isotopes in Fig.\ref{fig:bind_sn_n}. A
surprising property of the calculated energies is the almost constant,
very low neutron separation energy of the Fermi orbital from N=93 up
to the N=126 shell closure. This property is based on a large number
of orbitals being situated very close to the Fermi energy-- $^{176}$Sn
binds 34 neutrons within an energy interval of 0.84 MeV. For
comparison, the same number of neutrons in the same orbitals is spread
out over an energy interval of 3 MeV in the spectrum of
$^{208}$Pb. The large number of orbitals in a small energy interval
will make it very difficult to predict the neutron drip-line in the
Sn-isotopes by theoretical means and shows the need for experiments
with very exotic nuclides. The situation of orbitals is analogous to
the one in neutron rich Zirconium isotopes with A$>$122, which was shown
to lead to giant halos in relativistic Hartree-Boguliubov calculations
\cite{Men98}. This property should make the neutron-rich Sn nuclides a
very interesting system to study new modes of collectivity based on
the interaction with the continuum.

\section{Conclusion}

We have established a new parameterization for the Woods-Saxon potential.
Its six parameters are fitted to single-particle spectra around doubly
magic nuclides and experimental charge radii. We achieved a high quality
of description of the single-particle spectra observed around the
doubly-magic nuclei between $^{16}$O and $^{208}$Pb. The nuclear
properties such as radii, shell evolutions, magic numbers and driplines
are all shown to be well described considering limitations of the
Woods-Saxon approach. The isospin conservation by the nuclear part
of interaction and proper treatment of the two-body particle-core
kinematics are important for the success of the model. The parameter
set is intended to enable the use of the Woods-Saxon potential as
a basis for shell model calculations and a pathway to connect the
physics of bound and unbound nuclear states.

\section{Acknowledgements}

We would like to acknowledge the numerous discussions and contributions
by Prof. R.A.\ Wyss and Dr. A. Oros in a previous effort to determine
an optimized parameter set for the Woods-Saxon potential. This work
was supported partially by the NSF under contract No.0456463 and by
the DOE under contracts DE-FG02-02ER41220 and DE-FG02-92ER40750.

\end{document}